\documentclass[10pt,fleqn,aps,prd,nofootinbib,superscriptaddress,floatfix]{revtex4}
\usepackage[latin9]{inputenc}
\usepackage{helvet, mathpazo, amsmath, amssymb, color, textcomp, graphicx, amsfonts, graphics, epstopdf, slashed, multirow, adjustbox, lipsum, rotating, xcolor, wrapfig, epsfig,float}
\usepackage[unicode=true, bookmarks=false, breaklinks=false,pdfborder={0 0 1},backref=section,colorlinks=false]{hyperref}
\hypersetup{ colorlinks,bookmarksopen,bookmarksnumbered,linkcolor=blu,pdfstartview=FitH,urlcolor=rossos,citecolor=rossos}
\usepackage[justification=raggedright, margin=5mm,font=sf,small,labelfont=bf, format=plain]{caption}
\definecolor{rosso}{cmyk}{0,1,1,0.4}
\definecolor{rossos}{cmyk}{0,1,1,0.55}
\definecolor{rossoc}{cmyk}{0,1,1,0.2}
\definecolor{blu}{cmyk}{1,1,0,0.3}
\definecolor{blus}{cmyk}{1,1,0,0.6}
\definecolor{bluc}{cmyk}{1,1,0,0.1}
\definecolor{verde}{cmyk}{0.92,0,0.59,0.25}
\definecolor{verdec}{cmyk}{0.92,0,0.59,0.15}
\definecolor{verdes}{cmyk}{0.92,0,0.59,0.4}

\begin{document}
\title{\color{bluc}Dark Matter in a Singlet Extended Inert Higgs Doublet Model}

\author{Mohammed Omer Khojali $^{\href{0000-0002-0702-262X}{\includegraphics[width=2mm]{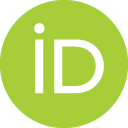}}} $}
\email{khogali11@gmail.com}
\affiliation{Department of Physics, University of Johannesburg, PO Box 524, Auckland Park 2006, South Africa.}
\affiliation{Department of Physics, University of Khartoum,\\ PO Box 321, Khartoum 11115, Sudan.}

\author{Ammar Abdalgabar $^{\href{0000-0001-9933-7313}{\includegraphics[width=2mm]{ORCID.png}}} $}
\email{amari@uhb.edu.sa}
\affiliation{Department of Physics, College of Science,\\University of Hafr Al Batin, Hafr Al Batin 39524, KSA.}
\affiliation{Department of Physics, Sudan University of Science and Technology, Khartoum 407, Sudan.}

\author{Amine Ahriche ${\href{https://orcid.org/0000-0003-0230-1774}{\includegraphics[width=2mm]{ORCID.png}}} $}
\email{ahriche@sharjah.ac.ae}
\affiliation{Department of Applied Physics and Astronomy, University of Sharjah,
P.O. Box 27272 Sharjah, UAE.}

\author{Alan S. Cornell $^{\href{https://orcid.org/0000-0003-1896-4628}{\includegraphics[width=2mm]{ORCID.png}}} $}
\email{acornell@uj.ac.za}
\affiliation{Department of Physics, University of Johannesburg, PO Box 524, Auckland Park 2006, South Africa.}

\begin{abstract}
In this work, we consider an extension of the Standard Model (SM) with an inert Higgs doublet and a real scalar singlet, in order to address problems around the origin of dark matter (DM). In this model, the lightest among the CP-odd and CP-even neutral inert components plays the role of a DM candidate, where the model parameters are subject to many theoretical and experimental constraints. These constraints include vacuum stability, perturbativity, LEP negative searches, electroweak precision tests, Higgs di-photon, Higgs invisible and Higgs undetermined decays, DM relic density and DM direct detection bounds. Using these constraints, we find that the allowed parameter space for these models is quite sizeable and could be explored in upcoming collider and astrophysical searches.
\end{abstract}

\maketitle

\section{Introduction}
The greatest achievement of the Large Hadron Collider (LHC) has been the discovery of the missing building block of the Standard Model (SM), the Higgs boson~\cite{ATLAS:2012yve,CMS:2012qbp}. This discovery opened a new era in particle physics, where whilst there has been no direct evidence of new physics beyond the SM, yet, there are many reasons to believe that new physics should be present at, or about, the \textrm{TeV} scale.

Even though the Higgs boson was successfully discovered, many questions remain. One such question is understanding the nature of Dark Matter (DM)~\cite{Bertone:2016nfn,Bertone:2004pz}. Though this is one of the great problems in the SM there are no strong clues to explain DM, even though it is needed to explain astrophysical and cosmological observations. This has motivated many people to consider extensions to the SM where one or more scalar fields are added. In this way, DM can be classified into thermal or non-thermal cases. In the thermal case, DM particles are in both thermal and chemical equilibrium with other particles in the thermal soup of a very early epoch of the universe. As the Universe expands and temperature decreases, the thermal DM candidate will freeze out and become detectable through relic density measurements. This is considered as the simplest scenario of DM candidates, known as a Weakly Interacting Massive Particle (WIMP)~\cite{LopezHonorez:2006gr,LopezHonorez:2010eeh}. The WIMP scenario is the most studied in the literature as it possesses many attractive properties and is also relevant to many types of DM searches. A WIMP could have a mass of the order of the electroweak weak scale, $m_{DM} \sim 10~\text{GeV}-1~\text{TeV} $, and has couplings to the SM fields of the order of the electroweak couplings~\cite{Chakraborti:2018aae}.

Several cosmological and astrophysical observations during the last decades have provided strong evidence for the existence of DM within the Universe. The amount of the cold dark matter (CDM) has been precisely measured by the Planck satellite mission as~\cite{Planck:2018vyg} 
\begin{equation}
\Omega_{\rm CDM}h^2 \,=\,0.120\pm\,0.0010,\label{omh}
\end{equation}
where the CDM content is estimated to comprise roughly $26\% $ of the overall energy within the Universe~\cite{Ade:2015xua}.

The phenomenology of DM has been investigated in various extensions of the SM, the simplest one being the inert Higgs doublet model (IDM), which has two SU(2) doublets in the scalar sector~\cite{Deshpande:1977rw}. The IDM was examined and constrained within the framework of LHC phenomenology, both in relation to the Higgs boson discovery and DM constraints. Moreover the model offered a rich phenomenology related to various aspects of DM, see Refs.~\cite{Deshpande:1977rw,LopezHonorez:2006gr,Gustafsson:2007pc}. The IDM, though theoretically well-motivated as a minimal consistent DM explanation, failed to adequately explain the existence of DM in the region of intermediate mass (100 - 500)~\textrm{GeV}. An extension of this scenario is to consider a type of model with extensions of the SM that may address the possible origin of DM. One such model is the singlet extended inert Higgs doublet model, where an inert Higgs doublet and a real scalar singlet is added the SM.

The IDM has a discrete $Z_2 $ symmetry under which the new inert scalar doublet is odd; whereas all the SM fields are even~\cite{Deshpande:1977rw}. More IDM phenomenology has been studied in Ref.~\cite{Gustafsson:2007pc}. This discrete $Z_2 $ symmetry has important consequences, such as the absence of flavor changing neutral currents at tree level~\cite{Glashow:1976nt}, and the absence of interactions with active fermions. This makes the lightest scalar among neutral CP-even and CP-odd components ( $H^{0}/A^{0} $) a good DM candidate. It has been shown that for scalar DM, which is lighter than $50~\textrm{GeV} $, the relic density is too large with respect to the observed value Eq.~(\ref{omh}) due to the suppressed annihilation cross section into SM light fermions, as implied by the null results from DM direct detection (DD) experiments. This situation can be avoided if there exists new mediators and other annihilation channels as in~\cite{Ahriche:2016cio}. In the IDM, the DM annihilation into $W^{+}W^{-} $ is too large for the mass range $140~\textrm{ GeV}<m_{H^{0}}\leq~550~\textrm{GeV} $, which makes the relic density too small. There exist three viable mass regions in the IDM:
\begin{itemize}
 \item[i)] around $m_{H^{0}}\simeq m_{h}/2 $, where the s-channel resonance Higgs boson exchange plays a key role~\cite{Borah:2017dqx}, 
 \item[ii)] around the $W $ gauge boson mass, that is driven by the annihilation into the three-body final state\\ $WW^{*}\rightarrow Wff' $~\cite{Honorez:2010re}, and 
 \item[iii)] $m_{H^{0}}\sim~\text{TeV} $ with scalar couplings of order unity~\cite{Choubey:2017hsq}.
\end{itemize}

In this work, we extend the IDM by a real singlet, that acquires a vacuum expectation value and mixes with the SM Higgs doublet. This induced mixing modifies all the interactions in the IDM, including the relevant one to the DM annihilation. Such a scenario has been previously studied in~\cite{DuttaBanik:2014iad,Bonilla:2014xba}, however, it deserves to be re-analysed due to the plethora of recent measurements made after 2015; such as the bounds from DM SI direct detection, the invisible, undetermined and di-photon Higgs decay channels and the Higgs strength modifier, in addition to the negative searches on heavy scalar resonances. As such, we impose an additional global $Z_2 $ symmetry in the model here, aside from the one that is responsible for DM stabilization, and which forbids terms like $S $, $S^2 $ and $\Phi_i^{\dagger} \Phi_i ~S $ in the Lagrangian. The absence of such terms does not change the phenomenology and the predictions of the model, but makes the parameter space smaller.

We shall, furthermore, consider loop effects, that are important when studying Higgs boson self-couplings. This requires the use of tools usually reserved for probing higher energies, chiefly the Renormalization Group Equations (RGEs)~\cite{Degrassi:2012ry}. Recall that RGEs provide a way by which partial explorations of the physics implications at a high energy scale are possible, as the theories at asymptotic energies may reveal new symmetries or other interesting properties that may lead to deeper insights into the physical content of the universe. In our model, there exist many additional scalar quartic interactions that involve the SM and BSM fields. In these situations, it is crucial to probe theoretical requirements like perturbativity and vacuum stability at high scales ~\cite{Jangid:2020qgo}.
 
Our paper is organised as follows: In Section~\ref{sec:Model} we present our model, including the field content and the renormalizable potential. In Section~\ref{sec:constr} we determine the theoretical and experimental constraints for various model parameters, such as unitarity, vacuum stability, perturbativity, electroweak precision tests, LHC constraints on the Higgs boson and heavy scalar masses, the Higgs boson strength modifier, the ratios $R_{\gamma\gamma} $ and $R_{\gamma Z} $ and the DM DD constraints. The DM relic density is given in Section~\ref{sec:DM}. Section~\ref{sec:NA} contains the numerical analysis and discussion. The RGEs are given in Section~\ref{sec:RGEs}. We conclude our paper in Section~\ref{sec:Conc}.

\section{Model\label{sec:Model}}

In this model, we extend the SM by an additional $SU(2)_L $ scalar inert doublet $\Phi $ and a real scalar singlet $S $, assigned with a global $ Z_2 \times Z_2^{\prime} $ symmetry, where the field quantum numbers and parities are shown in Table~\ref{table:QM}. The fact that the extra doublet is assumed to be odd under a discrete $Z_2 $ symmetry ensures the DM candidate's stability. The renormalizable potential has the form 
\begin{eqnarray}
V_{0} & =& m_1^2 {\cal H}^{\dagger}{\cal H} + m_2^2 \Phi^{\dagger}\Phi+\frac{1}{2}m_{s}^2 S^2 +\frac{1}{6}\lambda_1({\cal H}^{\dagger}{\cal H})^2+\frac{1}{6}\lambda_2(\Phi^{\dagger}\Phi)^2 +\frac{1}{24}\lambda_{S}S^{4}+\lambda_3 ({\cal H}^{\dagger}{\cal H} )(\Phi^{\dagger}\Phi )\nonumber \\
 && +\frac{1}{2}\omega_1S^2 {\cal H}^{\dagger}{\cal H}+\frac{1}{2}\omega_2S^2 \Phi^{\dagger}\Phi+\lambda_4 (\Phi^{\dagger}{\cal H} )({\cal H}^{\dagger}\Phi )+\frac{1}{2}\{\lambda_5 (\Phi^{\dagger}{\cal H} )^2 +h.c\}.\label{eq:V}
\end{eqnarray}

\begin{table}[t]
\centering
\begin{tabular}{|c|c|c|c|c|c|c|c|c|c|c|c|}
\hline 
Field &\ ${\cal H}$ &\ $\Phi$ &\ $S $ &\ $Q_L $ &\ $u_{R} $ &\ $d_{R} $ &\ $L_L $ &\ $\ell_{R} $ &\ $B_{\mu} $ &\ $W^{a}_{\mu} $ &\ $G_{\mu}^{i} $ \\ [0.5ex]
\hline
 $SU(2)\times U(1) $ & $(2,-1) $ & $(2,1) $ & $(1,0) $ & $(2,1/6) $ & $(1,-2/3) $ & $(1,1/3) $ & $(2,-1/2) $ & $(1,1) $ & $(1,0) $ &\ $(3,0) $ &\ $(1,0) $ \\
\hline
 $(Z_{2}, Z_{2}^{\prime}) $ & $(+, +) $ & $(-,+) $ & $(+,-) $ & $(+,+) $ & $(+,+) $ & $(+,+) $ & $(+,+) $ & $(+,+) $ & $(+,+) $ & $(+,+) $ & $(+,+) $ \\
\hline
 \end{tabular}
\caption{\it Quantum numbers and parities of the field content.}
\label{table:QM}
\end{table}

Here, ${\cal H}$ is the SM scalar doublet and the global symmetry $Z_2^{\prime} $ forbids terms in the scalar potential (\ref{eq:V}) such as $S $, $S^3 $, ${\cal H}^{\dagger}{\cal H} S $ and $\Phi^{\dagger}\Phi S $. However, this global $ Z_2^{\prime} $ symmetry is spontaneously broken, together with the electroweak symmetry, when the fields ${\cal H} $ and $S $ acquire vacuum expectation values (VEV) of
\begin{equation}
{\cal H} =\left(\begin{array}{c}
\chi^{+}\\
\frac{1}{\sqrt{2}}(\upsilon+h'+i\chi)
\end{array}\right)\,,~~~\Phi =\left(\begin{array}{c}
H^{+}\\
\frac{1}{\sqrt{2}}(H^{0}+iA^{0})
\end{array}\right)\,,~S=\upsilon_{s}+s,\label{eq:Sc}
\end{equation}
where $\upsilon=246.22\,\textrm{GeV} $ and $\upsilon_{s} $ denotes the singlet VEV. The tadpole conditions can be used to eliminate the parameters $m_1^2 $ and $m_{s}^2 $ in favor of the scalar VEVs. The inert scalar masses are given by
\begin{equation}
m^2_{H^{\pm}}=m^2_2+\frac{1}{2}\lambda_3 \upsilon^2+\frac{1}{2} \omega_2 \upsilon_s^2 ,~m^2_{H^0,A^0}=m^2_{H^{\pm}}+\frac{1}{2}(\lambda_4\pm\lambda_5)\upsilon^2,
\end{equation}
where the lightest among $H^{0} $ and $A^{0} $ plays the role of the DM candidate.

The mass eigenstates $h $ and $H $ are linear combinations of $h' $ and $s $ and can be written as 
\begin{equation}
h=s~\sin\alpha+h'~\cos\alpha\,,\,H=s~\cos\alpha-h'~\sin\alpha\,,
\end{equation}
where $\alpha $ is the mixing angle between $h $ and $H $. The $h' $- $s $ mixing due to the presence of the term of $\omega_1 $ in Eq. (\ref{eq:V}) leads to the squared mass matrix 
\begin{equation}
M^2 =\left(\begin{array}{cc}
\frac{1}{3}\,\lambda_1 \,\upsilon^2 & \omega_1 \,\upsilon\,\upsilon_{s}\\
\omega_1 \,\upsilon\,\upsilon_{s} & \frac{1}{3}\,\lambda_{S}\,\upsilon_{s}^2 
\end{array}\right),\label{massmatrix}
\end{equation}
which gives the eigenvalues and the mixing 
\begin{equation}
m_{1,2}^2 =\frac{1}{6}(\lambda_1 \,\upsilon^2 +\lambda_{S}\,\upsilon_{s}^2 \mp\sqrt{(\lambda_{S}\,\upsilon_{s}^2 -\lambda_1 \,\upsilon^2 )^2 +36\omega_1^2 \,\upsilon^2 \,\upsilon_{s}^2 }),\,t_{2\alpha}=\frac{6\omega_1 \,\upsilon\,\upsilon_{\phi}}{\lambda_2 \,\upsilon_{s}^2 -\lambda_1 \,\upsilon^2 }.\label{eq:mHeta}
\end{equation}

In this work, we denote $h=h_{125} $ as the SM-like observed Higgs boson and $H $ as an additional scalar that could be either heavier or lighter than the $h_{125} $ SM-like one. Then, in the case where the SM-like Higgs boson is the lighter one, we have
\begin{eqnarray}
m_{h,H}^2 &=&\frac{1}{6}\lambda_1 \upsilon^2 +\frac{1}{6}\lambda_{S}\upsilon_{s}^2 \mp\frac{1}{6}(\lambda_{S}\upsilon_{s}^2 -\lambda_1 \upsilon^2 )/c_{2\alpha},\notag \\
\lambda_1 &=&3\frac{m_{h}^2 \,c_{\alpha}^2 +m_{H}^2 \,s_{\alpha}^2 }{\upsilon^2 },\,\lambda_{S}=3\frac{m_{h}^2 \,s_{\alpha}^2 +m_{H}^2 \,c_{\alpha}^2 }{\upsilon^2 },\,\omega_1 =\frac{s_{2\alpha}}{2\upsilon\upsilon_{s}}\,(m_{H}^2 -m_{h}^2 ),
\end{eqnarray}
with $s_{X}=\sin X $ and $c_{X}=\cos X $. For the case of a heavier SM-like Higgs boson we get 
\begin{eqnarray}
m_{h,H}^2 &=&\frac{1}{6}\lambda_1 \upsilon^2 +\frac{1}{6}\lambda_{S}\upsilon_{s}^2 \pm\frac{1}{6}(\lambda_{S}\upsilon_{s}^2 -\lambda_1 \upsilon^2 )/c_{2\alpha},\notag \\
\lambda_1 &=&3\frac{m_{h}^2 \,s_{\alpha}^2 +m_{H}^2 \,c_{\alpha}^2 }{\upsilon^2 },\,\lambda_{S}=3\frac{m_{h}^2 \,c_{\alpha}^2 +m_{H}^2 \,s_{\alpha}^2 }{\upsilon^2 },\,\omega_1 =\frac{s_{2\alpha}}{2\upsilon\upsilon_{s}}\,(m_{h}^2 -m_{H}^2 ).
\end{eqnarray}

The model parameters are subject to many theoretical and experimental constraints, such as the vacuum stability, the perturbativity, the LEP negative searches (the search for $e^-e^+\rightarrow H Z $~\cite{OPAL:2002ifx}), the electroweak precision tests, the Higgs di-photon, the Higgs invisible and Higgs undetermined decays, the DM relic density and DM DD bounds, which we shall apply in the next section.

\section{Theoretical and experimental constraints\label{sec:constr}}

This model is subject to many theoretical and experimental constraints, as listed previously, where for each we find:

\begin{itemize}
\item {\bf Unitarity constraints}\\
The perturbative unitarity must be preserved in all the processes involving scalars and/or gauge bosons. At very high energy scales, the gauge bosons could be replaced by their Goldstone bosons, and therefore it would be easier to estimate the scattering amplitude matrix for only scalars of $S_{1}S_{2}\rightarrow S_{3}S_{4} $. This can be understood as, at high energies, the dominant contributions to these amplitudes are those mediated by the quartic couplings~\cite{Arhrib:2000is}. Here, $S_{1,2,3,4} $ can be any scalar degree of freedom in Eq. (\ref{eq:Sc}). The perturbative unitarity conditions are achieved only if the eigenvalues of the scattering amplitude matrix are smaller than $|\Lambda_{i}|<8\pi $~\cite{Cornwall:1974km}. Note that in any model, some discrete symmetries like the electric charge must be exact, while other symmetries like CP could be either exact or spontaneously/explicitly broken. In our model, both CP, the global $Z_{2}$ symmetry and the electric charge are exact symmetries, and therefore the full scattering amplitude matrix can be divided into sub-matrices according to the symmetry of the initial/final state $S_{1}S_{2}/S_{3}S_{4} $.

According to the initial state $S_1S_2 $ symmetry, whether it is CP even/odd, electrically neutral/charged and $Z_2 $ even/odd, we have six submatrices that are defined in the basis:
\begin{itemize}
 \item Neutral, CP-even and $Z_2 $ even: $\{hh,ss,\chi^0\chi^0,H^0H^0,A^0A^0,\chi^+\chi^-,H^+H^-\} $.
 \item Neutral, CP-even and $Z_2 $ odd: $\{hH^0,sH^0,\chi^0A^0,\chi^+H^-\} $.
 \item Neutral, CP-odd and $Z_2 $ even: $\{h\chi^0,s\chi^0,H^0A^0,\chi^+\chi^-,H^+H^-\} $.
 \item Neutral, CP-odd and $Z_2 $ odd: $\{h A^0,s A^0,\chi^0 H^0,\chi^+H^-\} $.
 \item Charged and $Z_2 $ even: $\{ \chi^{\pm}h,\chi^{\pm}s,\chi^{\pm}\chi^0,H^{\pm}H^0,H^{\pm}A^0 \} $.
 \item Charged and $Z_2 $ odd: $\{\chi^{\pm}H^0, \chi^{\pm} A^0, H^{\pm} h, H^{\pm} s, H^{\pm} \chi^0 \} $.
\end{itemize}

These submatrices are given in Appendix~\ref{eq:CP}.

\vspace{5mm}

\item {\bf Vacuum stability and perturbativity}\\
The quartic couplings of the scalar potential in Eq. (\ref{eq:V}) are subject to a number of constraints that ensure that the potential should be bounded from below and that the couplings remain perturbative. This can be guaranteed by the conditions
\begin{equation}
\lambda_{1,2,S},|\omega_{1,2}|, |\lambda_3|, |\lambda_{\pm}|\leq 4\pi , \label{eq:Unit}
\end{equation}
where $\lambda_{\pm}=\lambda_3+\lambda_4\pm \lambda_5 $. For the scalar potential to be bounded from below, the coefficient of the leading (quartic) term in any direction must be positive. This can be achieved via the conditions
\begin{equation}
\lambda_{1,2,S}>0,\,\,\text{and}\,\,
\begin{vmatrix}
\begin {array}{ccccccc}
\lambda_1 &\overline{\omega_1 }&\overline{\lambda_+ }&\frac{1}{3}\lambda_1 &\overline{\lambda_- }&\frac{1}{3}\lambda_1 &\overline{\lambda_3 }\\ \noalign{\medskip} \overline{\omega_1 }
&\lambda_{s}&\overline{\omega_2 }&\overline{\omega_1 }&\overline{\omega_2 }&\overline{\omega_1 }&\overline{\omega_2 }
\\ \noalign{\medskip} \overline{\lambda_+ }&\overline{\omega_2 }&\lambda_2 &\overline{\lambda_- }&\frac{1}{3}\lambda_2 &\overline{\lambda_3 }&\frac{1}{3}\lambda_2 \\ \noalign{\medskip} \frac{1}{3}\lambda_1 &\overline{\omega_1 }
&\overline{\lambda_- }&\lambda_1 &\overline{\lambda_+ }&\frac{1}{3}\lambda_1 &\overline{\lambda_3 }
\\ \noalign{\medskip} \overline{\lambda_- }&\overline{\omega_2 }&\frac{1}{3}\lambda_2 &\overline{\lambda_+ }&\lambda_2 &\overline{\lambda_3 }&\frac{1}{3}\lambda_2 \\ \noalign{\medskip}\frac{1}{3}\lambda_1 &\overline{\omega_1 }&\overline{\lambda_3 }&\frac{1}{3}\lambda_1 &\overline{\lambda_3 }&\frac{2}{3}\lambda_1 &\overline{\lambda_3+\lambda_4}
\\ \noalign{\medskip}\overline{\lambda_3 }&\overline{\omega_2 }&\frac{1}{3}\lambda_2 &\overline{\lambda_3 }&\frac{1}{3}\lambda_2 &\overline{\lambda_3+\lambda_4}&\frac{2}{3}\lambda_2 
\end{array} 
\label{eq:Vstab}
\end{vmatrix}
>0,
\end{equation}
where $\overline{X}=\text{Min}(X,0) $.

\vspace{5mm}

\item {\bf Electroweak precision tests}\\
The existence of an extra doublet, $\Phi $, and the mixing with a scalar singlet induces new contributions on the oblique parameters. While taking $\Delta U=0 $ in our analysis, the oblique parameters, $\varDelta T $ and $\varDelta S $ can be written as: 
\begin{eqnarray}
\varDelta T & = & \frac{1}{16\pi s_{\mathrm{w}}^2 m_{W}^2 }\left\{ F(m_{H^{0}}^2 ,m_{H^{\pm}}^2 )+F(m_{A^{0}}^2 ,m_{H^{\pm}}^2 )-F(m_{H^{0}}^2 ,m_{A^{0}}^2 )\right.\nonumber \\
 & & \left.+3s_{\alpha}^2 [F(m_{Z}^2 ,m_{H}^2 )-F(m_{Z}^2 ,m_{h}^2 )-F(m_{W}^2 ,m_{H}^2 )+F(m_{W}^2 ,m_{h}^2 )]\right\} ,\nonumber \\
\varDelta S & = & \frac{1}{24\pi}\left\{ (2s_{\mathrm{w}}^2 -1)^2 G(m_{H^{\pm}}^2 ,m_{H^{\pm}}^2 ,m_{Z}^2 )+G(m_{H^{0}}^2 ,m_{A^{0}}^2 ,m_{Z}^2 )+\log(m_{H^{0}}^2 m_{A^{0}}^2 /m_{H^{\pm}}^{4})\right.\nonumber \\
 & & \left.+s_{\alpha}^2 [\log(m_{H}^2 /m_{h}^2 )+G_{x}(m_{H}^2 ,m_{Z}^2 )-G_{x}(m_{h}^2 ,m_{Z}^2 )]\right\} ,\label{eq:oblique}
\end{eqnarray}
where $s_{\mathrm{W}}\equiv\sin~\theta_{W} $, with $\theta_{W} $ being the Weinberg mixing angle, and the functions $F $, $G $ and $G_{x} $ are loop integral functions which are given in the~\cite{Grimus:2008nb}.

\vspace{5mm}

\item {\bf LHC constraints on the Higgs boson}\\
Due to the fact that the Higgs couplings are modified in our model with respect to the SM, in addition to the existence of new particles, the Higgs total decay width and branching ratios are modified. Here, all couplings of the Higgs-gauge fields and Higgs-fermions are scaled by $c_{\beta} $ where the Higgs partial decay widths to the SM particles are scaled as $\Gamma(h\rightarrow X_{SM}\bar{X}_{SM})=c_{\beta}^2 \,\text{\ensuremath{\Gamma}}_{SM}(h\rightarrow X_{SM}\bar{X}_{SM}) $. In addition to the SM final states, the Higgs boson may decay into neutral inert scalars and into the new scalar, if kinematically allowed. Thus, the Higgs total decay width can be written as
\begin{equation}
\Gamma_{h}=\text{\ensuremath{\Gamma_{BSM}}}+c_{\beta}^2 \,\ensuremath{\Gamma}_{h}^{SM},\label{eq:Cases}
\end{equation}
where $\ensuremath{\Gamma}_{h}^{SM}=4.2~\mathrm{MeV} $ is the Higgs total decay width in the SM; $\text{\ensuremath{\Gamma_{BSM}}}=\text{\ensuremath{\Gamma_{inv}}+\ensuremath{\Gamma_{und}}} $. $\ensuremath{\Gamma_{inv}=\varGamma\,(h\rightarrow H^{0}H^{0})+\varGamma\,(h\rightarrow A^{0}A^{0})} $ and $\ensuremath{\Gamma_{und}=\varGamma\,(h\rightarrow HH)} $ represents the invisible and the undetermined Higgs partial decay widths, respectively. These partial widths are given by
\begin{equation}
\varGamma\,(h\rightarrow SS)=\Theta(m_{h}-2m_{S})\frac{\lambda_{hSS}^2 }{32\pi m_{h}}\sqrt{1-4\frac{m_{S}^2 }{m_{h}^2 }},\label{Gund}
\end{equation}
with $S=H^{0},\,A^{0} $, and $\lambda_{hH^0H^0,hA^0A^0} = (\lambda_3+\lambda_4\pm\lambda_5)\,c_{\alpha}\,v-\omega_2 \,v_{s}\,s_{\alpha} $. Considering the recent measurements by ATLAS on the invisible~\cite{ATLAS:2020kdi} and undetermined~\cite{Aad:2019mbh} channels, we have 
\begin{equation}
\mathcal{B}_{inv}<0.11,\quad\text{and}\quad\mathcal{B}_{und}<0.22.\label{eq:Bh}
\end{equation}

As will be seen in Section~\ref{sec:NA}, the invisible Higgs constraint becomes irrelevant due to the tension from the combination between the constraints from the direct detection dark matter cross section and the di-photon Higgs decay. From the recent measurements, an upper bound on the total Higgs boson decay width has been established give $\Gamma_h < 14.4~\textrm{MeV} $ at the 68\% CL~\cite{ATLAS:2018jym}, however, we will consider a more conservative value by looking at the off-shell Higgs boson production in the final state $h\rightarrow Z Z^{*}\rightarrow\ell\ell\nu\nu $, $\Gamma_{h}=3.2^{+2.4}_{-1.7}~\textrm{MeV} $~\cite{CMS:2021ziv}.

\vspace{5mm}

\item {\bf The Higgs strength modifier}\\
The signal strength modifier measures the experimental quantity for the combined production and decay, and is defined as the ratio of the measured Higgs boson decay rate to its SM prediction. An effective method of studying the coupling of the Higgs boson is to study its deviation from the SM expectations. One has to consider that for a given search channel the signal strength modifier can be approximated to identify the Higgs boson production cross section times the decay branching ratios, normalized to the SM one ~\cite{Arcadi:2019lka}. For the $h \rightarrow XX $ decay channel, it is necessary to use the narrow width approximation
\begin{equation}
 \mu_{XX} \simeq \frac{\sigma(pp\rightarrow h \rightarrow XX)}{ \sigma(pp\rightarrow h \rightarrow XX)|_{SM}}\simeq \frac{\sigma(pp\rightarrow h)\times \mathcal{B}(h \rightarrow XX)}{\sigma(pp\rightarrow h)|_{SM}\times \mathcal{B}(h \rightarrow XX)|_{SM}}.
\end{equation}

As mentioned above, because of the Higgs mixing, the couplings of the observed Higgs boson to SM fermions and gauge bosons are modified with respect to the SM, by $c_{\beta} $. Therefore, the signal strength modifier can be written as:
\begin{equation}\label{mu}
 \mu_{XX}= c_{\beta}^2 (1-\mathcal{B}_{BSM}),
\end{equation}
where $\mathcal{B}_{BSM} = \mathcal{B}_{inv} $ (in the case of $m_H > m_h/2 $) and $\mathcal{B}_{BSM} = \mathcal{B}_{inv}+\mathcal{B}_{und} $ (in the case of $m_{H}<m_h/2 $).

As a result, a substantial limit on the mixing angle ${\beta} $ can be derived from the measurement of $\mu_{XX} $ of the $h_{125} $ Higgs boson. The limitations in Eq.~(\ref{mu}) complement those on the unusual Higgs boson decays shown in Eq.~(\ref{eq:Bh}). The total Higgs signal strength modifier was reported by ATLAS and CMS measurements at $\sqrt{s}= 7+8 $\,TeV to be $\mu_{XX}\ge 0.89 $ at a $95\% $ CL~\cite{Arcadi:2019lka}, which indicate $s_{\beta}^2 \le 0.11 $ in the absence of invisible and
undetermined Higgs decay $(\mathcal{B}_{BSM}=0) $.

\vspace{5mm}

\item {\bf LHC constraints of the heavy scalar $H $}

For the heavy CP even scalar (for $m_H>m_h $) decays into SM final states, di-Higgs or through other invisible or undetermined channels, there are different search types:\\
{\bf a)} The search for a heavy CP even resonance in the channels with a pair leptons, jets or gauge bosons $pp\rightarrow h \rightarrow \ell\ell,jj,VV $. In this search we consider the recent measurements by ATLAS at 13~\textrm{TeV} with 139~fb $^{-1} $ $pp\rightarrow h \rightarrow \tau\tau $~\cite{ATLAS:2020zms}, and $pp\rightarrow h \rightarrow ZZ $, through the channels $\ell\ell\ell\ell $ and $\ell\ell\nu\nu $~\cite{ATLAS:2020tlo}, as well as the CMS analysis at 13~\textrm{TeV} with 137~fb $^{-1} $ $pp\rightarrow h \rightarrow WW $~\cite{CMS:2021klu}.\\
{\bf b)} The second search type is via a resonant di-Higgs production $pp\rightarrow h \rightarrow HH $, where here we consider the recent ATLAS combination, which includes the analyses at 13~\textrm{TeV} with 139~fb $^{-1} $ through the channels $HH\rightarrow b\overline{b}\tau\tau $, $HH\rightarrow b\overline{b}b\overline{b} $ and $HH\rightarrow b\overline{b}\gamma\gamma $~\cite{ATLAS:2021jki}.\\
{\bf c)} The undetermined Higgs boson decay is different from the invisible one at colliders as the light scalar can be seen at detectors via the decay to light fermions $H\rightarrow f\bar{f} $. This decay does not match the known SM channels, but the undetermined signal $h\rightarrow HH\rightarrow f_1 \bar{f_1 }f_2 \bar{f_2 } $ can be probed.

In order to implement the first and second types of searches, one has to estimate the cross sections at the LHC at 13 \textrm{TeV}, as:
\begin{align}
\sigma (pp\rightarrow H \rightarrow XX) & = s_{\alpha}^2 \sigma^{SM}(pp\rightarrow H) \mathcal{B}(H \rightarrow XX),\label{eq:XS}
\end{align}
with $X=\ell,j,V, h $ and $\sigma^{SM}(pp\rightarrow H) $ is estimated in~\cite{LHC}.

The partial SM decay widths in our model can be written as
\begin{equation}
\Gamma(H \rightarrow XX) = s_{\beta}^2 \Gamma^{SM}(H \rightarrow XX)= s_{\beta}^2 \mathcal{B}^{SM}(H \rightarrow XX) \Gamma_{H}^{SM},
\end{equation}
and the total decay width is given by
\begin{equation}
\Gamma_{tot}(H) = s_{\beta}^2 \Gamma_{H}^{SM}+\Gamma(H \rightarrow hh)+\Gamma(H \rightarrow A^{0}A^{0})+\Gamma(H \rightarrow H^{0}H^{0})+\Gamma(H \rightarrow H^{+}H^{-}),\label{eq:HY}
\end{equation}

where the values of $\mathcal{B}^{SM}(H \rightarrow XX) $ and $\Gamma_{H}^{SM} $ are given in~\cite{LHC}. The partial decay widths in Eq.~(\ref{eq:HY}) are given by
\begin{equation}
\varGamma\,(H\rightarrow YY)=\Theta(m_{H}-2m_{Y})\frac{q_Y\lambda_{HYY}^2 }{32\pi m_{H}}\sqrt{1-4\frac{m_{Y}^2 }{m_{H}^2 }},
\end{equation}
with $Y=h,\,H^{0},\,A^{0},\,H^{\pm} $ and $q_h=q_{H^0}=q_{A^0}=1,\,q_{H^{\pm}}=2 $, where the corresponding scalar triple couplings $\lambda_{HYY} $ are given by
\begin{align}
\lambda_{Hhh} & = \lambda_1 \,c_{\alpha}^2 \,s_{\alpha}\,v+\lambda_{s}\,c_{\alpha}\,s_{\alpha}^2 \,v_{s}+\omega_1 (s_{\alpha}^{3}\,v\,-2\,c_{\alpha}^2 \,s_{\alpha}\,v\,+c_{\alpha}^{3}\,v_{s}-2\,c_{\alpha}\,s_{\alpha}^2 \,v_{s}),\nonumber\\
\lambda_{HH^0H^0} & = (\lambda_3+\lambda_4+\lambda_5)\,s_{\alpha}\,v+\omega_2 \,v_{s}\,c_{\alpha},\nonumber \\
\lambda_{HA^0A^0} & = (\lambda_3+\lambda_4-\lambda_5)\,s_{\alpha}\,v+\omega_2 \,v_{s}\,c_{\alpha},\nonumber \\
\lambda_{HH^+H^-} & = s_{\alpha}\,v\,\lambda_3+c_{\alpha}\,v_{s}\,\omega_2 .
\end{align} 

\vspace{5mm}

\item {\bf The ratio $R_{\gamma\gamma} $ and $R_{\gamma Z} $}\\
The decays of the SM Higgs boson into $\gamma V $ ( $V=\gamma,Z $) occurs via loops which are mediated by $W $ bosons as well as heavy charged fermion loops~\cite{Djouadi:2005gi}. In the model under consideration, these decays are modified due to the extra contribution of the inert charged scalar, which modifies the ratio $R_{\gamma V}=\mathcal{B}(h\rightarrow \gamma V)/\mathcal{B}^{SM}(h\rightarrow \gamma V) $ as
\begin{eqnarray}
R_{\gamma\gamma} & = & (1- \mathcal{B}_{BSM}) \left\lvert 1 +\frac{\lambda_{hH^{+}H^{-}}\,v}{2m^2_{H^{\pm}}c_{\alpha}} \frac{A_{0}^{\gamma\gamma}\left(\tau_{H^{\pm}}\right)}{A_1^{\gamma\gamma}\left(\tau_W\right)+\frac{4}{3}A_{1/2}^{\gamma\gamma}\left(\tau_t\right)} \right\rvert^2 ,\nonumber\\
R_{\gamma Z} & = & (1-\mathcal{B}_{BSM}) \left\lvert 1-\frac{1-2s_{w}^2 }{c_{w}} \frac{\lambda_{hH^{+}H^{-}}\,v}{2m^2_{H^{\pm}}c_{\alpha}} 
\frac{A_{0}^{\gamma Z}\left(\tau_{H^{\pm}},\lambda_{H^{\pm}}\right)}{c_{w} A_1^{\gamma Z}\left(\tau_W, \lambda_W\right)+\frac{6-16s_{w}^2 }{3c_{w}}A_{1/2}^{\gamma Z}\left(\tau_t,\lambda_t\right)}
\right\rvert^2 ,
\end{eqnarray}
with $\mathcal{B}_{BSM}=\mathcal{B}_{inv}-\mathcal{B}_{unv} $ and $\tau_X=\frac{m_{h}^2 }{4m_X^2 } $ and $\lambda_X=\frac{m_Z^2 }{4m_X^2 } $. The loop functions $A_{i}^{\gamma V}(x), (i=0,1,1/2) $ are given in~\cite{Djouadi:2005gi}.

\vspace{5mm}

\item {\bf DM direct detection constraints}\\

The DD of elastic nucleon-DM scattering has provided the most rigorous constraints on DM mass and interactions in a large number of conventional DM models.

\begin{figure}[t]
\centering
\includegraphics[width=0.3\textwidth]{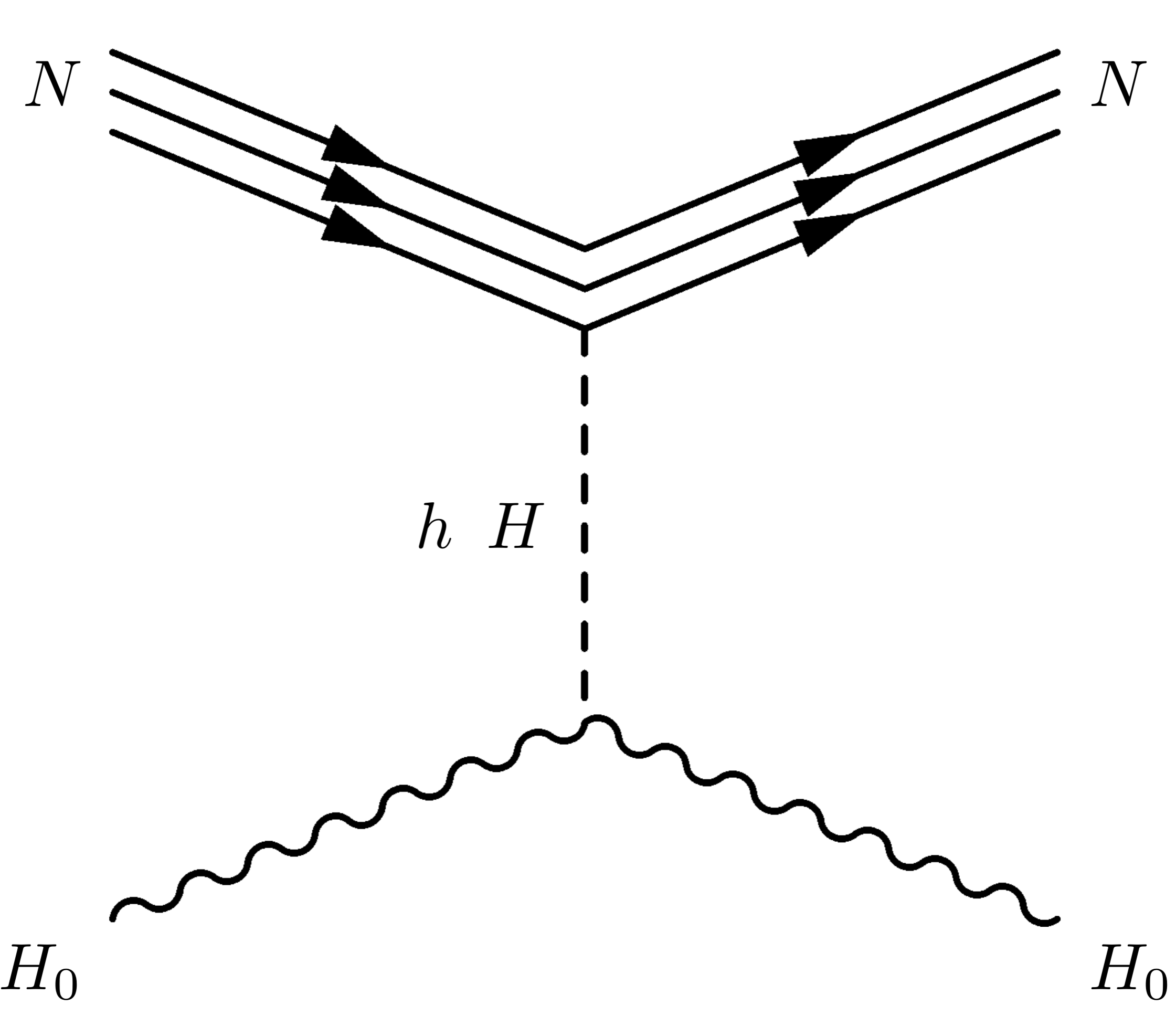}
\caption{Feynman diagram of the DD cross section for the scattering of the DM candidate, $H_0 $, off a nucleon.}
\label{Feyn:fig}
\end{figure}
According the Feynman diagram in Fig.~\ref{Feyn:fig}, the DD cross section for the scattering of the DM candidate ( $H^{0} $ for example) in this model off a nucleon is given by~\cite{He:2008qm}
\begin{equation}
\sigma_{det}=\frac{g_{hNN}^2 m_{N}^2 }{4\pi(m_{N}+m_{DM})^2 }\left[\frac{\lambda_{hH^{0}H^{0}}c_{\alpha}}{m_{h}^2 }-\frac{\lambda_{HH^{0}H^{0}}s_{\alpha}}{m_{H}^2 }\right]^2 ,\label{Sigdet}
\end{equation}
where $m_{N} $ is the nucleon mass in the chiral limit, $\lambda_{hH^{0}H^{0},HH^{0}H^{0}} $ are the scalar triple couplings, and $g_{hNN} $ is the effective Higgs-nucleon coupling. The Higgs-nucleon coupling is estimated using a heavy baryon chiral perturbation theory to be $g_{hNN}\simeq(1.07\pm0.26)\times10^{-3}~ $~\cite{Alarcon:2011zs}. Note though that lattice calculations give smaller values~\cite{QCDSF:2011mup}. From Eq.~(\ref{Sigdet}), allowed values for the DD cross section can be achieved by a proper choice of the couplings $\lambda_{hH^0H^0,HH^0H^0} $, the mixing $s_{\alpha} $ and the mass $m_H $. Unlike the IDM case, small DD cross sections can be obtained only for very suppressed values of $\lambda_L=\lambda_3+\lambda_4\pm\lambda_5 $.

\end{itemize}

\section{Dark Matter relic density\label{sec:DM}}

According to the WIMP scenario, the DM relic abundance can be estimated
by solving the so-called Zel'dovich-Okun-Pikelner-Lee-Weinberg equation
(ZOPLW)~\cite{ZOPLW} that describes the DM number density $n_{H^{0}} $
evolution; 
\begin{equation}
\frac{dn_{H^{0}}}{dt}+3\,H\,n_{H^{0}}=-\,\left<\sigma\upsilon\right>\big(n_{H^{0}}^2 -(n_{H^{0}}^{eq})^2 \big),\label{ZOPLW}
\end{equation}
where $\left<\sigma\upsilon\right> $ is the thermally averaged annihilation
cross section of the DM particles times their relative velocity, $H $
is the Hubble expansion parameter and $n_{H^{0}}^{eq} $ is the DM
equilibrium number density. In cases where next-to-DM particles (here
the CP-odd inert $A^{0} $) masses are close to the DM mass, the co-annihilation
effect becomes important and needs to be included in Eq. (\ref{ZOPLW}).
In order to do, the thermally averaged annihilation cross section
 $\left<\sigma\upsilon\right> $ should be replaced by an effective
one at the temperature $T=m_{H^{0}}/x $; 
\begin{equation}
\langle\upsilon\sigma_{eff}(x)\rangle=\frac{x}{8m_{H^{0}}}\frac{\sum_{i,j}g_{i}g_{j}\int_{(m_{i}+m_{j})^2 }^{\infty}\,\frac{ds}{\sqrt{s}}\,K_1 \big(x\frac{\sqrt{s}}{m_{H^{0}}}\big)\lambda^2 (s,m_{i}^2 ,m_{j}^2 )\langle\upsilon\sigma\rangle_{ij}(s)}{\Big[\sum_{i}g_{i}m_{i}^2 K_2 \Big(x\frac{m_{i}}{m_{H^{0}}}\Big)\Big]^2 },\label{eq:Sigeff}
\end{equation}
with $i,j=H^{0},~A^{0} $, $K_{1,2} $ are the modified Bessel functions,
 $\langle\upsilon\sigma\rangle_{ij}(s) $ is the annihilation cross
section of the process $ij\rightarrow SM $ at the CM energy $\sqrt{s} $;
and $\lambda(x,y,z)=\sqrt{x^2 +y^2 +z^2 -2(xy+xz+yz)} $.

To solve numerically the ZOPLW equation and estimate the freeze-out
and relic density, we use MadDM~\cite{Ambrogi:2018jqj}, where we
use FeynRules~\cite{Alloul:2013bka} to generate the required UFO
files.

\section{Numerical Analysis and Discussion\label{sec:NA}}

\begin{figure}[t]
\includegraphics[width=0.48\textwidth]{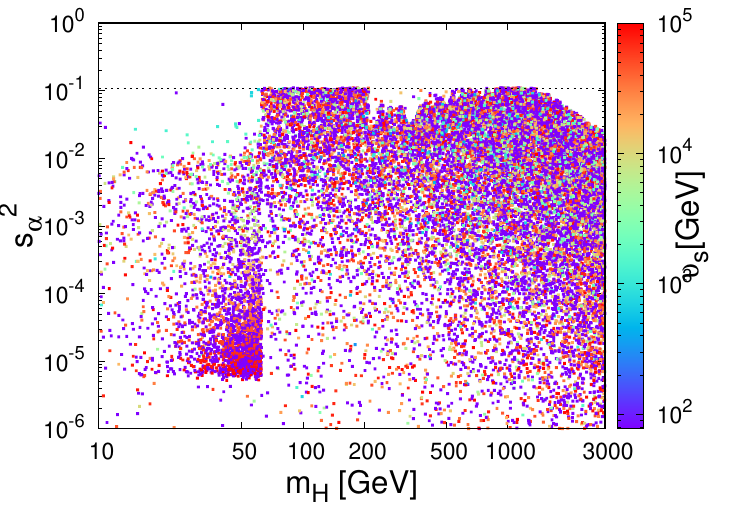} \includegraphics[width=0.48\textwidth]{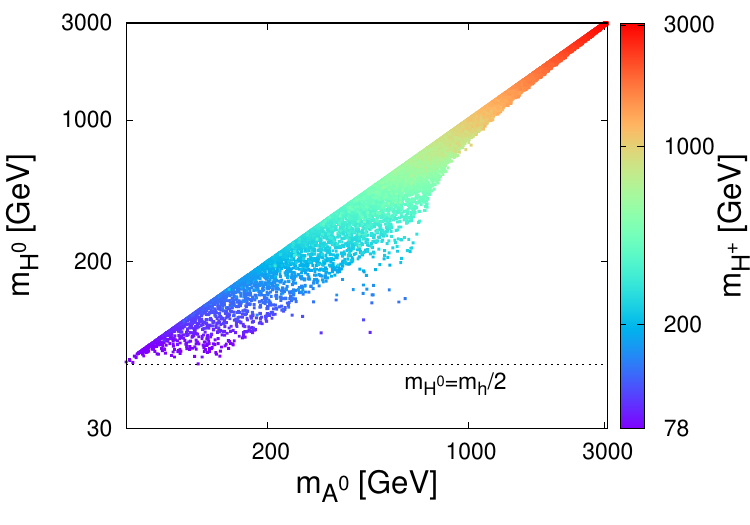}\\
 \includegraphics[width=0.48\textwidth]{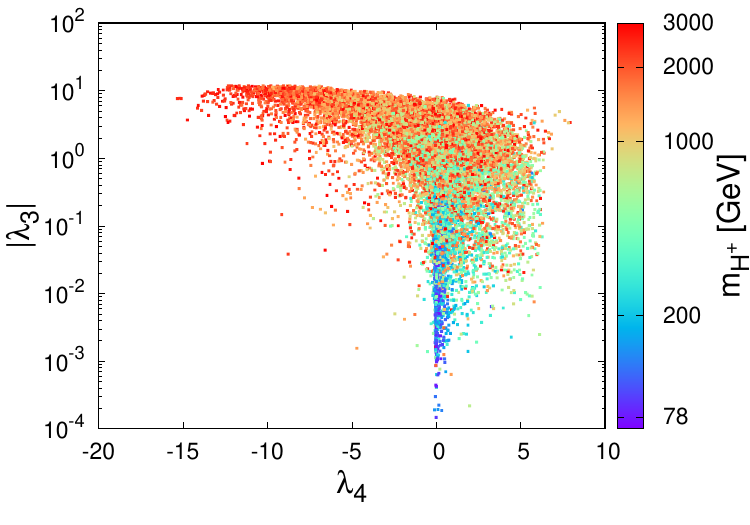} \includegraphics[width=0.48\textwidth]{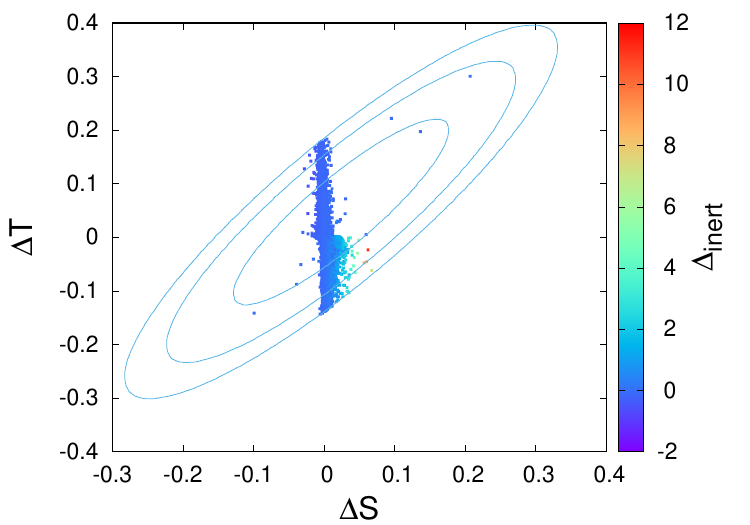}
\caption{Top-left: the mixing $s_{\alpha}^2 $ versus the extra Higgs boson mass $m_{H} $, where the palette shows the singlet VEV $v_{s} $.Top-right:
the inert masses $m_{H^{0},A^{0},H^{\pm}} $ for the 20k considered
BPs. Bottom-left: The coupling $|\lambda_3 | $ versus $\lambda_4 $,
where the palette shows the mass $m_{H^{\pm}} $. Bottom-right: the
oblique parameters constraint due to the oblique parameters $\Delta T $
and $\Delta S $, with the palette showing the inert relative mass difference
 $\Delta_{inert}=(m_{H^{0}}^2 +m_{A^{0}}^2 -2m_{H^{\pm}}^2 )/2m_{H^{\pm}}^2 $.}
\label{constraints} 
\end{figure}

In our numerical analysis, we make a random scan over the following ranges for the model free parameters: 
\begin{align}
m_{h}<m_{H}<3~\textrm{TeV},~113.5~\textrm{GeV}<m_{H^{\pm}}<3~\textrm{TeV},~1~\textrm{GeV}<m_{H^{0},A^{0}}<3~\textrm{TeV},\nonumber \\
~10~\textrm{GeV}<\upsilon_{s}<10~\textrm{TeV},~\max(|\omega_2 |,~\lambda_2 ,~|\lambda_3 |)<4\pi,\label{eq:PS}
\end{align}
where we take into account all the above mentioned constraints except the relic density. Note also that all the 20k benchmark points (BPs) used in this analysis are in agreement with the experimental bounds from PandaX-4T 2021~\cite{PandaX-4T:2021bab}. In Fig.~\ref{constraints}, we present the allowed ranges of the model free parameters that satisfy all the theoretical and experimental constraints listed above for 20k BPs.

From the top-left panel in Fig.~\ref{constraints}, one learns that the new extra CP-even scalar could be either heavier or lighter than the SM-like Higgs boson for all possible values of the mixing and the singlet VEV. In the top-right panel, we display the values of the inert masses $m_{H^{0},A^{0},H^{\pm}} $. Clearly, the mass range difference could not be larger than 500 GeV due to the perturbativity conditions on the couplings $\lambda $'s. In the bottom-left panel, we show $|\lambda_3 | $ versus $\lambda_4 $ for the values of $m_{H^{\pm}} $ shown in the palette. In this case, for most of the BPs with light charged scalar mass, the coupling $\lambda_3 $ must take values well below 0.1, due to the constraints from the di-photon Higgs decay $R_{\gamma\gamma} $. However, for the heavier charged scalar, this constraint is less severe and therefore the $\lambda_3 $ could be large $\mathcal{O}(1) $. $\lambda_4 $ can take large values as the conditions on $\lambda_{\pm} $ in Eq. (\ref{eq:Unit}) is fulfilled. In the bottom-right panel, we show the new physics contributions to the oblique parameter constraints $\Delta S $ and $\Delta T $, where the different ellipses correspond to $68\% $, $95\% $ and $99\% $ CL intervals obtained from the precise measurement of various observables. Note that most of the BPs correspond to negative values of $\Delta_{inert}=(m_{H^{0}}^2 +m_{A^{0}}^2 -2m_{H^{\pm}}^2 )/2m_{H^{\pm}}^2 $, and therefore negative $\lambda_4 $.

\begin{figure}[t]
\centering \includegraphics[width=0.49\textwidth]{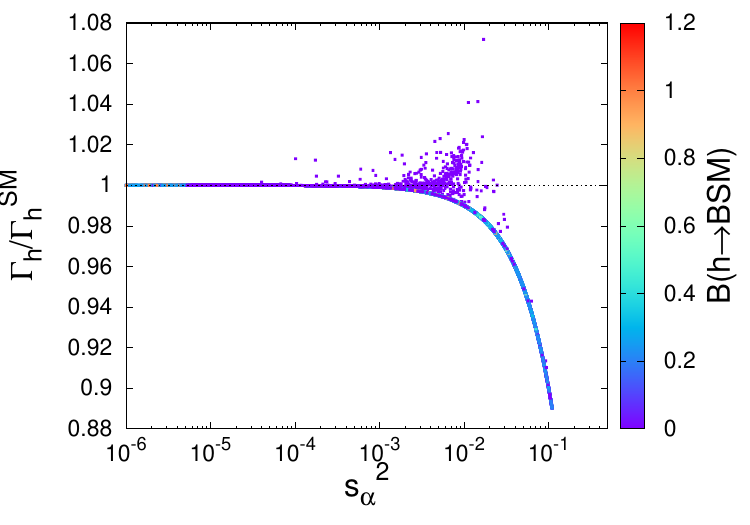}~\includegraphics[width=0.49\textwidth]{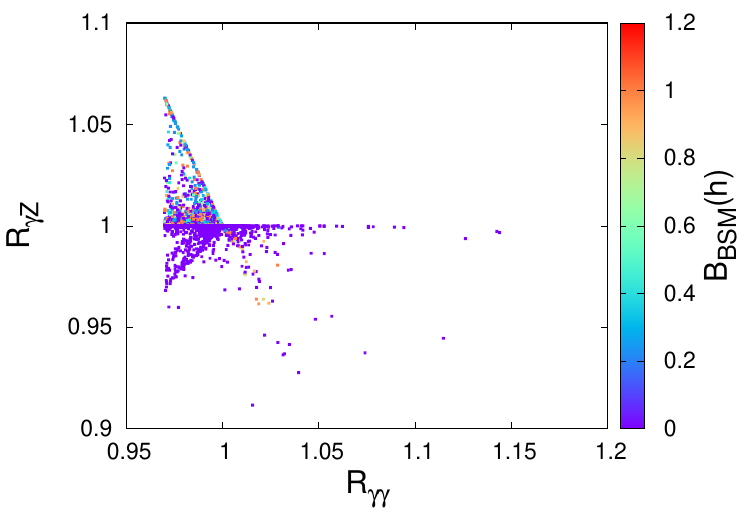}\\
\includegraphics[width=0.49\textwidth]{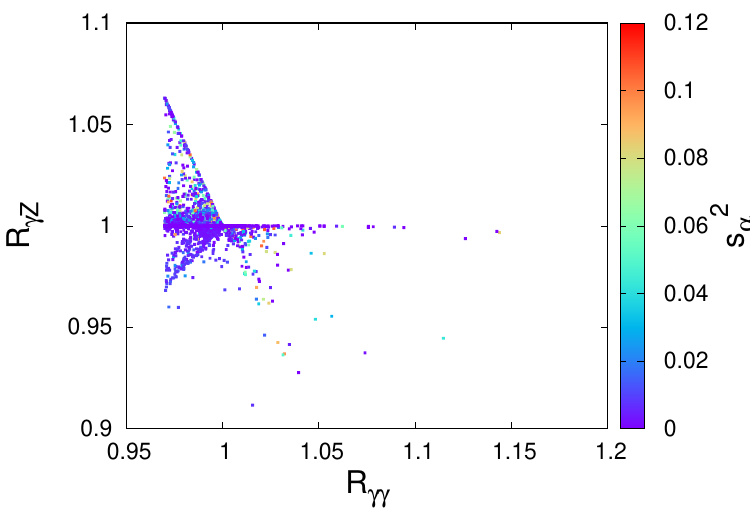} \caption{Left: the Higgs total decay width scaled by its SM value versus the mixing $s_{\alpha}^2 $, where the palette shows the BSM (invisible+undetermined) branching ratio. For the middle and right panels, the ratios $R_{\gamma\gamma,\gamma Z} $ where the palette shows the BSM branching ratio (middle) and the mixing $s_{\alpha}^2 $ (right).}
\label{higgs} 
\end{figure}

\begin{figure}[t]
\includegraphics[width=0.49\textwidth]{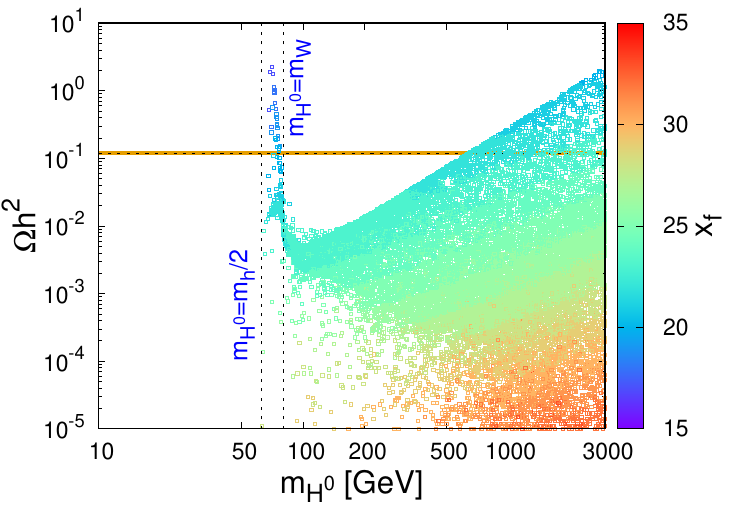}~\includegraphics[width=0.49\textwidth]{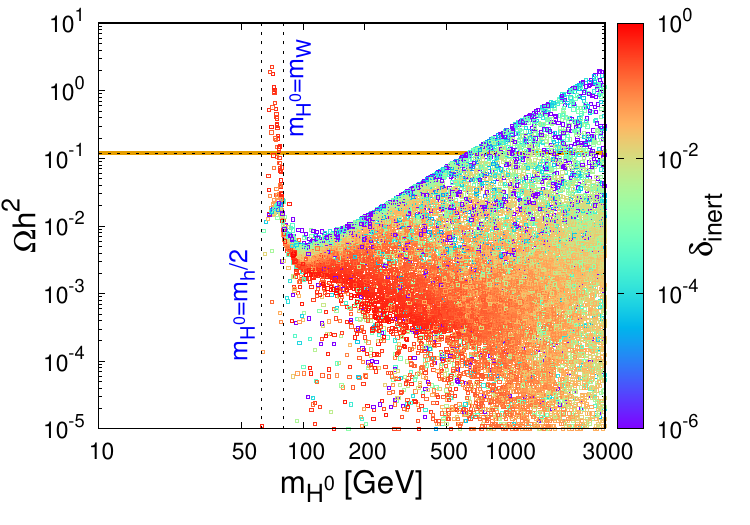}\\
\includegraphics[width=0.49\textwidth]{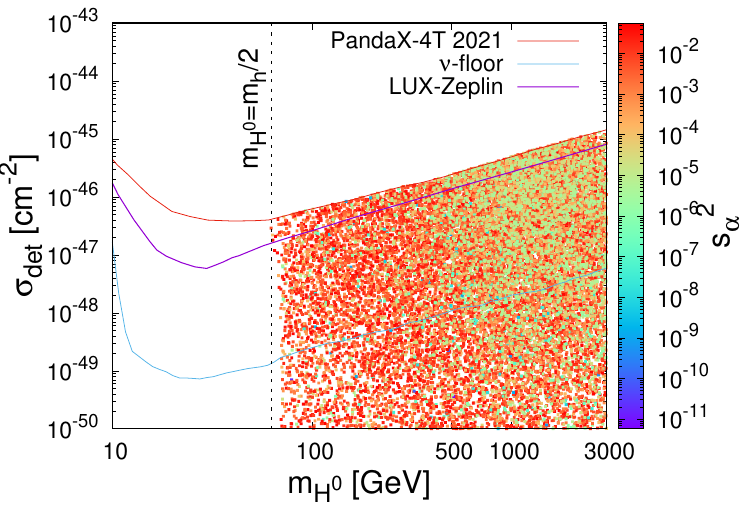}
\caption{The DM relic density $\Omega h^2 $ versus the DM mass, where the palette shows the freeze-out temperature $x_{f} $ (left) and the mass splitting $\delta_{inert}=(m^2_{A^0}-m^2_{H^0})/(m^2_{A^0}+m^2_{H^0}) $ (right). The tight orange band represents the observed relic density value (\ref{omh}), the DD cross section $\sigma_{det} $ as a function of the DM mass $m_{H^{0}} $, and the current bounds on the DD cross section are from PandaX-4T 2021~\cite{PandaX-4T:2021bab}, LUX-ZEPLIN~\cite{LUX-ZEPLIN:2022qhg} ; and $\nu $-floor~\cite{Billard:2013qya}, with the palette showing $s_{\alpha}^2 $ (Bottom).}\label{DM}
\end{figure}

In Fig.~\ref{higgs}, we show the observables that are relevant to the SM-like Higgs boson such as the ratio $R_{\gamma\gamma} $, the Higgs invisible and undetermined branching ratios; and the total decay width. The left panel in Fig.~\ref{higgs} shows for most of the BPs the Higgs total decay width scales like $s_{\alpha}^2 $ due to the absence of invisible ( $h\rightarrow H^{0}H^{0},A^{0}A^{0} $), and/or undetermined ( $h\rightarrow HH $) decay channels. Indeed, the BPs that lead to the total Higgs boson decay width being larger than the SM value correspond to the allowed undetermined decay channel ( $h\rightarrow HH $) since the invisible one is not allowed, as will be seen in Fig.~\ref{DM}. In the middle and right panels, the ratio $R_{\gamma Z} $ lies between 0.92 and 1.07 for the experimentally allowed values of $R_{\gamma\gamma}=1.09\pm0.12 $~\cite{ATLAS:2017ovn}. From the palettes, one reads that for relatively large mixing and large $B_{BSM}(h) $ the ratio $R_{\gamma\gamma} $ is always reduced, while where the ratio $R_{\gamma Z} $ could be enhanced up to 7\%.

\begin{figure}[t]
\includegraphics[width=0.48\textwidth]{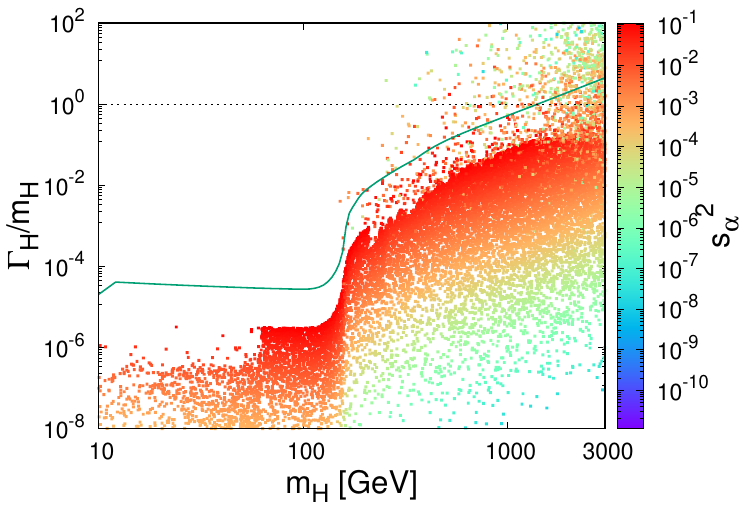}~\includegraphics[width=0.48\textwidth]{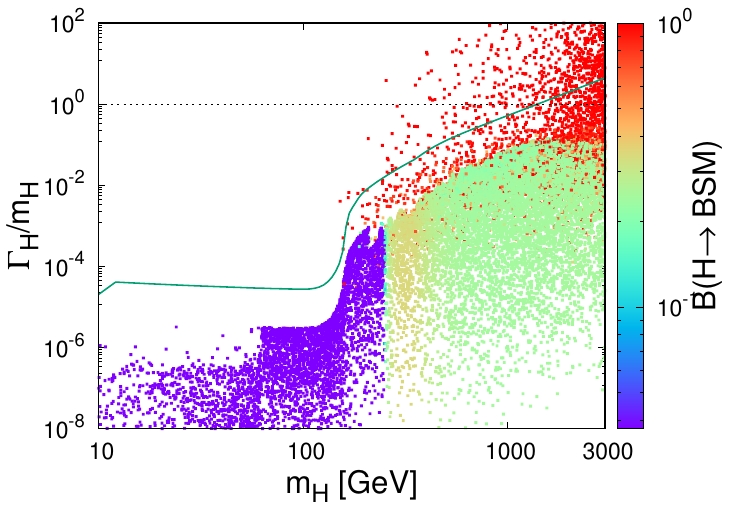}
\caption{The total decay width of the new CP-even scalar $H $ scaled by its mass, where the palette shows the mixing (left) and its BSM branching fraction (right). Here, the green curve corresponds to the total decay width $\Gamma_{H} $ for the SM interactions, i.e., for $s_{\alpha}=1 $ and $\mathcal{B}_{BSM}=0 $.}
\label{HH1} 
\end{figure}

\begin{figure}[t]
 \includegraphics[width=0.48\textwidth]{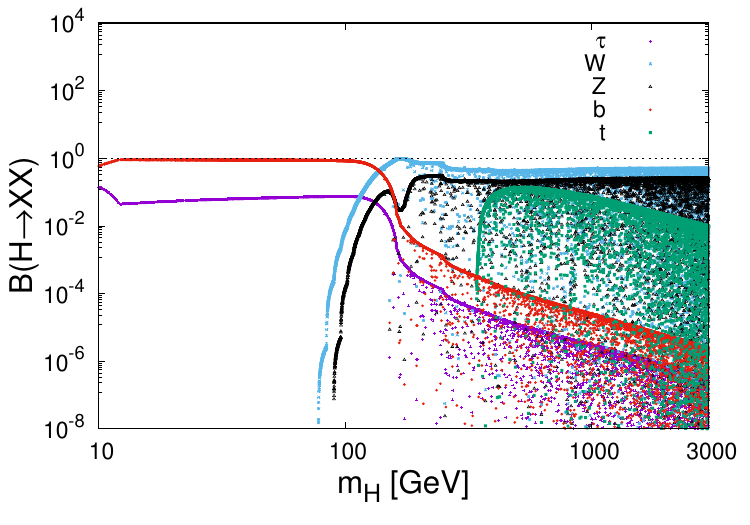} \includegraphics[width=0.48\textwidth]{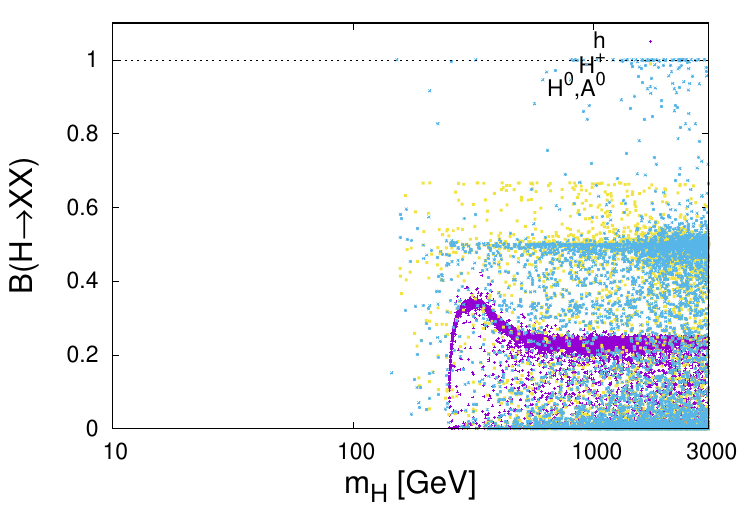}
\caption{The branching ratios of the heavy scalar ( $H $) into the SM final states $(X=\tau,W,Z,b,t) $ (left) and into the non-SM ones $(X=h,H^{\pm},H^{0},A^{0}) $, versus its mass $m_{H} $.}
\label{HH2} 
\end{figure}

In Fig.~\ref{HH1} and Fig.~\ref{HH2}, we present different properties of the new CP-even scalar, such as its total decay width and branching fractions. From Fig.~\ref{HH1}, one notices that for most of the BPs, the total decay width is one order or magnitude smaller than the SM values due to the factor $s_{\alpha}^2 \le 0.11 $, as can be read from the palette in the top-left panel. For the BPs with $m_{H}<m_{h} $, the BSM channels $H\rightarrow H^{0}H^{0},A^{0}A^{0} $ are not allowed, while for the BPs with $m_{H}>m_{h} $, these channels and/or the channel $H\rightarrow hh $ could be dominant. One has to notice that the BPs with dominant $H\rightarrow hh $ are very interesting since they could be the subject of many experimental searches, especially in the channels $pp\rightarrow h\rightarrow HH\rightarrow b\bar{b}b\bar{b},b\bar{b}\tau\tau,\tau\tau\tau\tau $~\cite{ATLAS:2021ldb}. Fig.~\ref{HH2} shows that the new scalar $H $ decay is dominated by the BSM channels $(hh,H^{0}H^{0},A^{0}A^{0},H^{\pm}H^{\pm}) $ for large $m_{H} $ values, and by the channel $b\bar{b} $ for smaller $m_{H} $ values.

In Fig.~\ref{DM}, we show the DM relic density as a function of the DM mass $m_{H^{0}} $, where the palette shows the freeze-out parameter $x_{f}=m_{H^{0}}/T_{f} $ (left) and the mass splitting $\delta_{inert}=(m^2_{A^0}-m^2_{H^0})/(m^2_{A^0}+m^2_{H^0}) $ (middle). In the right panel, we present the DM DD cross section vs the DM mass compared with the experimental bounds from PandaX-4T 2021~\cite{PandaX-4T:2021bab} and LUX-ZEPLIN~\cite{LUX-ZEPLIN:2022qhg}, where the palette shows the mixing $s^2_{\alpha} $. As mentioned previously, all the BPs used in this analysis are in agreement with the experimental bounds from PandaX-4T 2021~\cite{PandaX-4T:2021bab}.

From Fig.~\ref{DM}, one learns that the DM allowed mass range is between $m_h/2<m_{H^{0}}<m_W $ and $m_{H^{0}}>620~\textrm{GeV} $, which corresponds to the freeze-out parameter $x_{f}\sim 20-24 $. Although small values of the mass splitting $\Delta_{inert}=(m_{A^0}-m_{H^0})/(m_{A^0}+m_{H^0}) $ can make the co-annihilation effect important, it could not change the shape of the BPs in Fig.~\ref{DM}-left. It is clear also that DM with a mass $m_{H^{0}}<m_h/2 $ is excluded due the combination of many constraints. This can be understood from the fact that when imposing the DD and the di-photon Higgs decay bounds together, the Higgs invisible decay becomes dominant, which excludes any viable DM with $m_{H^{0}}<m_h/2 $. Fig.~\ref{DM}-right shows that this model DM can accommodate any future DD bounds up to the neutrino floor for different scalar mixings and new scalar masses.

\begin{figure}[t]
\centering
\includegraphics[width=0.48\textwidth]{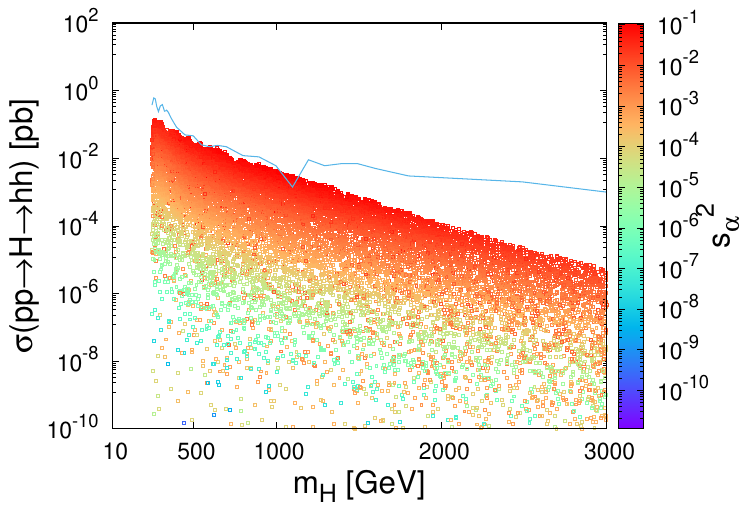}
\includegraphics[width=0.48\textwidth]{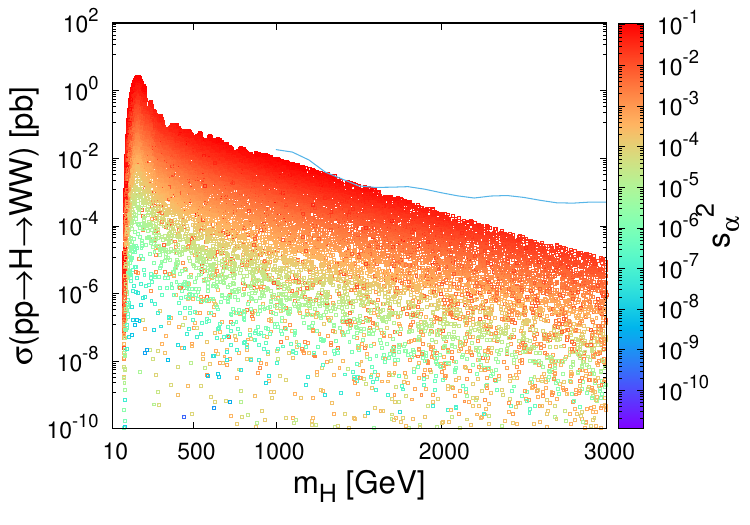}\\
\includegraphics[width=0.48\textwidth]{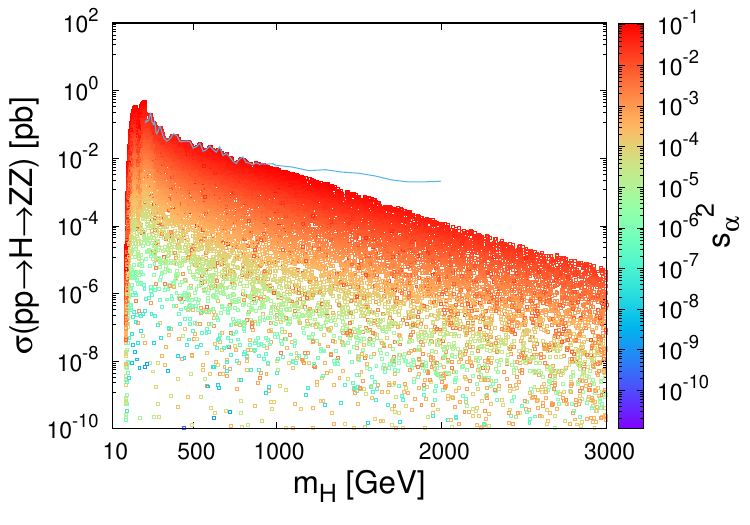}
\includegraphics[width=0.48\textwidth]{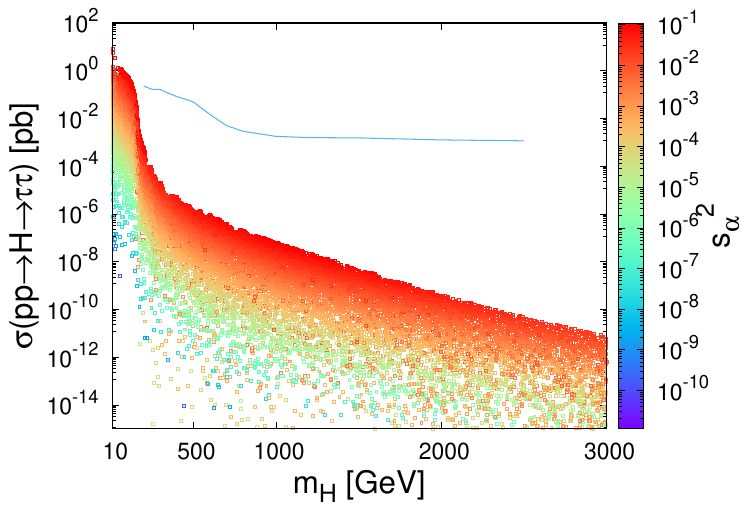}
\caption{The cross sections at the LHC for 13 \textrm{TeV} for the processes $\sigma(pp\rightarrow H \rightarrow hh,WW,ZZ,\tau\tau) $ versus the extra CP-even scalar mass, where the palette shows the mixing $s_{\alpha}^2 $. The blue line represents the experimental bounds from the ATLAS $139~fb^{-1} $ $hh\rightarrow b\bar{b}\tau\tau, b\bar{b}b\bar{b},b\bar{b}\gamma\gamma $~\cite{ATLAS:2021jki}, $pp\rightarrow\eta\rightarrow\tau\tau $~\cite{ATLAS:2020zms}, $pp\rightarrow\eta\rightarrow ZZ $~\cite{ATLAS:2020tlo} and CMS $137~fb^{-1} $ $pp\rightarrow\eta\rightarrow WW $~\cite{CMS:2021klu}.}\label{Cross}
\end{figure}

At the LHC, the new CP-even scalar can be produced and detected through its decay into different final states, in addition, the resonant di-Higgs channel $pp\rightarrow H \rightarrow hh $. As mentioned earlier, we consider here the recent measurements by ATLAS at 13~\textrm{TeV} with 139~fb $^{-1} $ $pp\rightarrow h \rightarrow \tau\tau $~\cite{ATLAS:2020zms}, and $pp\rightarrow h \rightarrow ZZ $, through the channels $\ell\ell\ell\ell $ and $\ell\ell\nu\nu $~\cite{ATLAS:2020tlo}, as well as the CMS analysis at 13~\textrm{TeV} with 137~fb $^{-1} $ $pp\rightarrow h \rightarrow WW $~\cite{CMS:2021klu}. While, for the resonant di-Higgs production we consider the final states $HH\rightarrow b\overline{b}\tau\tau $, $HH\rightarrow b\overline{b}b\overline{b} $ and $HH\rightarrow b\overline{b}\gamma\gamma $~\cite{ATLAS:2021jki}; obtained by ATLAS at 13~\textrm{TeV} with 139~fb $^{-1} $ integrated luminosity. Then, in Fig.~\ref{Cross}, we estimate the cross sections at the LHC for 13 \textrm{TeV} as defined in Eq. (\ref{eq:XS}) for the processes $\sigma(pp\rightarrow H \rightarrow hh, WW, ZZ, \tau\tau) $, versus the heavy Higgs boson mass $m_{H} $, and the mixing angle $s_{\alpha}^2 $ in the palette, compared to their relevant ATLAS and CMS upper bounds~\cite{ATLAS:2020zms,ATLAS:2020tlo,CMS:2021klu,ATLAS:2021jki}.

From Fig.~\ref{Cross}, one notices that most of the BPs are not excluded by these bounds, however, within the coming analyses with more integrated luminosity the bounding curves in Fig.~\ref{Cross} will become lower, and therefore some of the BPs would be excluded. A significant part of the parameter space that corresponds to the mass range $125~\textrm{GeV}<m_{H}<250~\textrm{GeV} $ will be probed if this mass range will be considered in the future analyses of ATLAS and CMS.

\section{Perturbativity and Vacuum Stability at High Scale}\label{sec:RGEs}

This model involves many more interactions than the SM and the IDM, and this will induce many contributions to the quantum corrections of the quartic couplings. This ensures that the perturbativity and vacuum stability conditions are different from the SM and IDM cases, and hence, this point needs to be investigated. The constraints that arise from vacuum stability, and that the couplings at higher scales remain perturbative, can be determined by the renormalisation group equation (RGE) of the gauge, Yukawa and quartic couplings. By neglecting all the Yukawa couplings, except for $y_{t} $ in what follows, we will use the $\beta $ functions at one- and two-loop level (as listed in Appendix~\ref{eq:RGE}) to check whether the conditions for the vacuum stability, perturbativity and unitarity are fulfilled at higher scales, such as $\Lambda=10^3\,\textrm{TeV},\,10^5\,\textrm{TeV},\,10^7\,\textrm{TeV} $. These $\beta $-functions are estimated using the module $\mathsf{CalcRGEs} $ of SARAH~\cite{Staub:2013tta}.

\begin{figure}[t]
\centering
\includegraphics[width=0.48\textwidth]{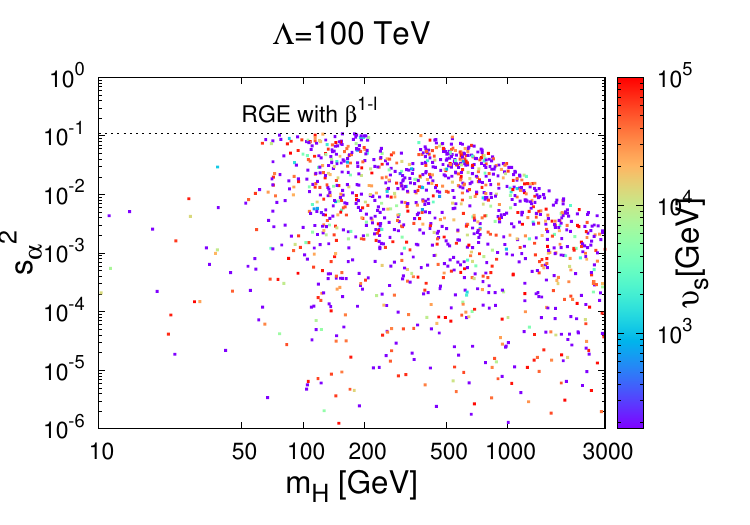}~\includegraphics[width=0.48\textwidth]{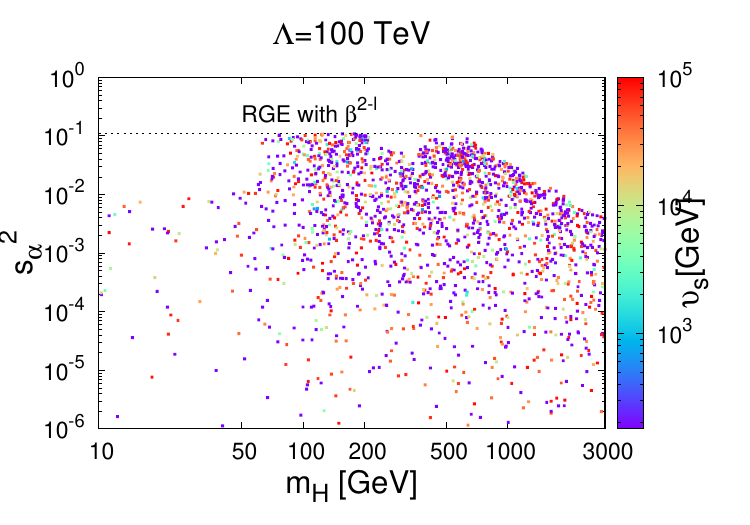}
\caption{The BPs as shown in Fig.~\ref{constraints}-top-left, that fulfill the perturbativity and the vacuum stability at $\Lambda=10^3~\textrm{TeV} $.}\label{1000TeV}
\end{figure}

We consider the 20k BPs used in Fig.~\ref{constraints}, estimating the running of the couplings at higher scales, and then show in Fig.~\ref{1000TeV} only the BPs that are in agreement with the perturbativity and the vacuum stability conditions at $\Lambda=10^3~\textrm{TeV} $. For instance, among the 20k PBs shown, these conditions on the couplings estimated at $\Lambda=10^3~\textrm{TeV} $ using one-loop (two-loop) $\beta $-functions, lead to only 1296 (1911) viable points as shown in Fig.~\ref{1000TeV}. Similar analysis in~\cite{Plascencia:2015xwa}, where they studied the high scale validity of the CSI IDM for $m_H>500 $ \textrm{GeV} up to one loop level, found fewer points survive to the Planck scale as we observe in our RGEs running up to one loop level. However, once the two loop level comes into play, we find that the regions in the parameter space which are viable up to the Planck scale are significantly enhanced in the scalar IDM case presented here.

\begin{figure}[t]
\centering
\includegraphics[width=0.48\textwidth]{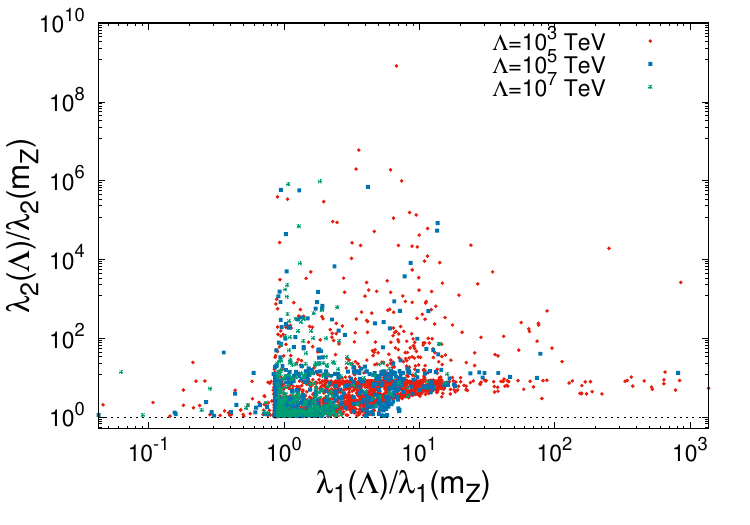}~\includegraphics[width=0.48\textwidth]{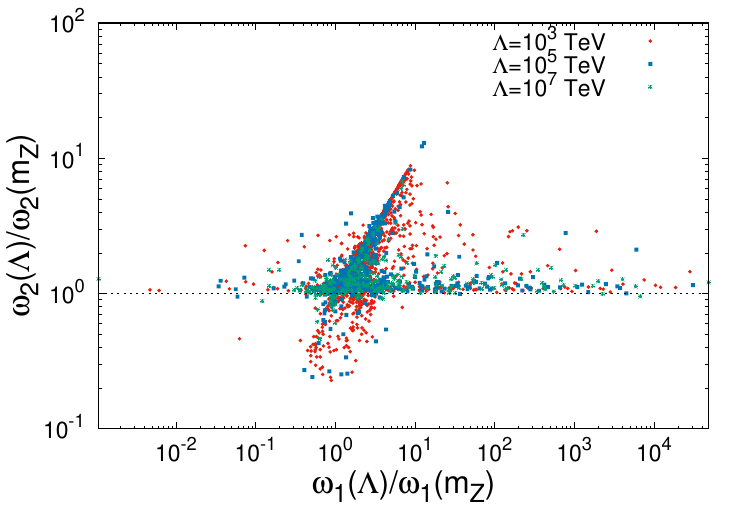}
\caption{The quartic couplings enhancement for a running scale using two-loop RGE at different scales.}\label{LamW}
\end{figure}

In Fig.~\ref{LamW}, we show the quartic couplings enhancement at $\Lambda=10^3\,\textrm{TeV},\,10^5\,\textrm{TeV},\,10^7\,\textrm{TeV} $ using two-loop RGEs. Indeed, only BPs that are in agreement with the perturbativity and the vacuum stability are considered. These conditions allow only 1911, 640 and 328 BPs at the scales $\Lambda=10^3\,\textrm{TeV},\,10^5\,\textrm{TeV},\,10^7\,\textrm{TeV} $, respectively. The evolution of the couplings $\lambda_{1,2} $ and $\omega_{1,2} $ at these scales are given in Fig.~\ref{LamW}.

The enhancement or the reduction of the value of quartic couplings can be significant according to the considered BPs, therefore a lot of BPs have been excluded at high scale from the scan by including these RGE/stability considerations. 

\section{Conclusion\label{sec:Conc}}

In this paper, we considered the inert Higgs Doublet Model extended by a real scalar singlet. In this setup, the model accommodates two CP-even eigenstates that have SM-like couplings to the SM fermions and gauge bosons, one of them $h=h_{125} $ is identified as the SM Higgs boson with the measured mass $m_h=125~\textrm{GeV} $ and the other scalar $H $ could be lighter or heavier the $h_{125} $ Higgs boson. Here, the DM candidate could be the lightest among the CP-even $H^0 $ and the CP-odd $A^0 $ scalars, where we adopted $H^0 $ to be the DM particle. In order to investigate the possibility of a DM candidate from this model, we performed a detailed numerical study to determine different regions of the parameter space that is consistent with current theoretical and experimental constraints, such as vacuum stability, perturbativity, unitarity, LEP negative searches, electroweak precision tests, experimental bounds on DM DD, the observed DM relic density, as well as the constraints from the Higgs boson decay.

Within this model, the DM relic density is well below the value measured by the Planck collaboration for the majority of the BPs, except for the mass ranges $m_h/2<m_{H^{0}}<m_W $ and $m_{H^{0}}>620~\textrm{GeV} $, which corresponds to the freeze-out parameter $x_{f}\sim 20-24 $. The co-annihilation effect can be important for small values of the mass splitting $\Delta_{inert}=(m_{A^0}-m_{H^0})/(m_{A^0}+m_{H^0}) $, but it does not change the shape of the BPs as shown in Fig.~\ref{DM}-left. In addition, DM lighter than $m_h/2 $ is excluded due to the combination of different constraints: DM
DD, $h\rightarrow\gamma \gamma $ and the Higgs invisible decay. This makes the model within the reach of high energy collider experiments. Also in this model we have shown that the DM is within the reach of future DD bounds, up to the neutrino floor for different scalar mixings and new scalar masses.

From our numerical analysis, we found that for most of the BPs considered, the total decay width is one order of magnitude smaller than the SM values, where for the BPs that led to the total Higgs boson decay width being larger than the SM value, these corresponded to the allowed undetermined decay channel ( $h\rightarrow HH $). Since the new scalar particle $H $ can be produced and detected at the LHC, the ATLAS and CMS negative searches on heavy resonances could be very useful to constraint the parameter space. We found that the current bounds can barely constraint the parameter space, but within the upcoming analyses with more integrated luminosity, the parameters space can be significantly probed through the channels $pp\rightarrow H \rightarrow ZZ,WW,b\bar{b}, \tau\tau $.

We have also considered the conditions of perturbativity and vacuum stability by running of the quartic scalar and gauge couplings at high scales using RGEs, but these conditions were not fulfilled at higher scales for a large number of the BPs, and so these can be excluded the conditions of vacuum stability, perturbativity and unitarity at higher scales. 
However, there are still remained a number of viable BPs, which should be testable at the next generation of experimental results. This means that this model may require UV completion at a scale between $10^5~\textrm{GeV} $ and $10^7~\textrm{GeV} $. 

\section*{Acknowledgments}
 MOK was supported by the GES. The work of AA is supported by the University of Sharjah under the research projects No 21021430100 "\textit{Extended Higgs Sectors at Colliders: Constraints \& Predictions}" and No 21021430107 "\textit{Hunting for New Physics at Colliders}". ASC is partially supported by the National Research Foundation South Africa.

\appendix

\section{Perturbative unitarity matrices}\label{eq:CP}

The neutral CP-even and $Z_2 $ even matrix in the basis $\{hh,ss,H^0H^0,\chi^0\chi^0,A^0A^0,\chi^+\chi^-,H^+H^-\} $ is given by
\begin{equation}
 \left(
\begin{array}{ccccccc}
 \lambda_1 & \omega_1 & \lambda_+ & \lambda_1 & \lambda_- & \frac{1}{3} \lambda_1 & \lambda_3 \\
 \omega_1 & \lambda_s & \omega_2 & \omega_1 & \omega_2 & \omega_1 & \omega_2 \\
 \lambda_+ & \omega_2 & \lambda_2 & \lambda_+ & \frac{1}{3}\lambda_2 & \lambda_3 & \frac{1}{3}\lambda_2 \\
 \lambda_1 & \omega_1 & \lambda_+ & \lambda_1 &\lambda_- & \frac{1}{3}\lambda_1 & \lambda_3 \\
 \lambda_- & \omega_2 & \frac{1}{3}\lambda_2 &\lambda_- & \lambda_2 & \lambda_3 & \frac{1}{3} \lambda_2 \\
 \frac{1}{3} \lambda_1 & \omega_1 &\lambda_3 & \frac{1}{3}\lambda_1 & \lambda_3 & \frac{2}{3} \lambda_1 & \lambda_3+\lambda_4 \\
 \lambda_3 & \omega_2 & \frac{1}{3}\lambda_2 &\lambda_3 & \frac{1}{3} \lambda_2 & \lambda_3+\lambda_4 & \frac{2}{3} \lambda_1 \\
\end{array}
\right) 
\end{equation}

 The neutral, CP-even and $Z_2 $ odd matrix in the basis $\{hH^0,sH^0,\chi^0A^0,\chi^+H^-\} $ is given by
\begin{equation}
\left(
\begin{array}{cccc}
 \lambda_+ & 0 & 0 & \frac{1}{2}(\lambda_4+\lambda_5) \\
 0 & \omega_2 & 0 & 0 \\
 0 & 0 & \lambda_- & \frac{i}{2}(\lambda_4-\lambda_5) \\
 \frac{1}{2}(\lambda_4+\lambda_5) & 0 & \frac{i}{2}(\lambda_4-\lambda_5) & \lambda_3+\lambda_4 \\
\end{array}
\right)
 \end{equation}
 
This matrix has the eigenvalues $\omega_2$, other eigenvalues will be calculated numerically.

The neutral CP-odd and $Z_2 $ even matrix in the basis $\{h\chi^0,s\chi^0,H^0A^0,\chi^+\chi^-,H^+H^- \} $ is given by
\begin{equation}
\left(
\begin{array}{ccccc}
 \lambda_1 & 0 & 0 & \frac{1}{3}\lambda_1 & \lambda_3 \\
 0 & \omega_1 & 0 & 0 & 0 \\
 0 & 0 & \frac{1}{3}\lambda_2 & 0 & 0 \\
 \frac{1}{3}\lambda_1 & 0 & 0 & \frac{2}{3}\lambda_1 & \lambda_3+\lambda_4 \\
 \lambda_3 & 0 & 0 & \lambda_3+\lambda_4 & \frac{2}{3}\lambda_2 \\
\end{array}
\right)
\end{equation}

This matrix has the eigenvalues $\omega_1 $ and $\frac{1}{3}\lambda_2 $, while the remaining 3 eigenvalues should be estimated numerically.

The neutral, CP-odd and $Z_2 $ odd matrix in the basis $\{h A^0,s A^0,\chi^0 H^0,\chi^+H^-\} $ is given by 

\begin{equation}
 \left(
\begin{array}{cccc}
 \lambda_- & 0 & 0 & \frac{i}{2}(\lambda_4-\lambda_5) \\
 0 & \omega_2 & 0 & 0 \\
 0 & 0 & \lambda_+ & \frac{1}{2}(\lambda_4+\lambda_5) \\
 -\frac{i}{2}(\lambda_4-\lambda_5) & 0 & \frac{1}{2}(\lambda_4+\lambda_5) & \lambda_3+\lambda_4 \\
\end{array}
\right)
\end{equation}

The charged and $Z_2 $ even matrix in the basis $\{ \chi^{\pm}h,\chi^{\pm}s,\chi^{\pm}\chi^0,H^{\pm}H^0,H^{\pm}A^0 \} $is given by
\begin{equation}
\left(
\begin{array}{ccccc}
 \frac{1}{3}\lambda_1 & 0 & \frac{1}{3}\lambda_1 & \frac{1}{2}(\lambda_4+\lambda_5) & \frac{i}{2}(\lambda_4-\lambda_5) \\
 0 & \omega_1 & 0 & 0 & 0 \\
 \frac{1}{3}\lambda_1 & 0 & \frac{1}{3}\lambda_1 & \frac{1}{2}(\lambda_4+\lambda_5) & \frac{i}{2}(\lambda_4-\lambda_5) \\
 \frac{1}{2}(\lambda_4+\lambda_5) & 0 & \frac{1}{2}(\lambda_4+\lambda_5) & \frac{1}{3}\lambda_2 & 0 \\
 -\frac{i}{2}(\lambda_4-\lambda_5) & 0 & -\frac{i}{2}(\lambda_4-\lambda_5) & 0 & \frac{1}{3}\lambda_2 \\
\end{array}
\right),
\end{equation}
which leads to the eigenvalues $0,\omega_1,\frac{1}{3}\lambda_2,\frac{1}{6} \left(2 \lambda_1+\lambda_2\pm\sqrt{(2 \lambda_1+ \lambda_2)^2+36 \lambda_4^2+36 \lambda_5^2}\right)$. 

 The charged and $Z_2 $odd matrix in the basis $\{\chi^{\pm}H^0,\chi^{\pm}A^0,H^{\pm}h,H^{\pm}s,H^{\pm}\chi^0 \} $ is given by
\begin{equation}
 \left(
\begin{array}{ccccc}
 \lambda_3 & 0 & \frac{1}{2}(\lambda_4+\lambda_5) & 0 & \frac{1}{2}(\lambda_4+\lambda_5) \\
 0 & \lambda_3 & \frac{i}{2}(\lambda_4-\lambda_5) & 0 & \frac{i}{2}(\lambda_4-\lambda_5) \\
 \frac{1}{2}(\lambda_4+\lambda_5) & -\frac{i}{2}(\lambda_4-\lambda_5) & \lambda_3 & 0 & \lambda_3 \\
 0 & 0 & 0 & \omega_2 & 0 \\
 \frac{1}{2}(\lambda_4+\lambda_5) & -\frac{i}{2}(\lambda_4-\lambda_5) & \lambda_3 & 0 & \lambda_3 \\
\end{array}
\right)
\end{equation}

This has eigenvalues $0,\omega_2,\lambda_3,\frac{1}{2} \left(3 \lambda_3 \pm \sqrt{\lambda_3^2+4 \lambda_4^2+4 \lambda_5^2}\right) $.

\section{One and Two-loop Renormailzation Group Equations }\label{eq:RGE}

The $\beta $ functions are given at one-loop level by:
\begin{align}
\beta_{g_1 } & =\frac{21}{5}g_1^{3},\,\beta_{g_2 }=-3g_2^{3},\,\beta_{g_3 }=-7g_3^{3},\nonumber \\
\beta_{\lambda_1 } & =\frac{81}{100}g_1^{4}+\frac{27}{10}g_1^2 g_2^2 +\frac{27}{4}g_2^{4}+12\lambda_3^2 +12\lambda_3 \lambda_4 +6\lambda_4^2 +6\lambda_5^2 +3\omega_1^2 -36y_{t}^{4}+4\lambda_1 \Big(\lambda_1 -\frac{9}{5}g_1^2 -9g_2^2 +12y_{t}^2 \Big),\nonumber \\
\beta_{\lambda_2 } & =12\lambda_3^2 +12\lambda_3 \lambda_4 +3\omega_2^2 +4\lambda_2^2 +6\lambda_4^2 +6\lambda_5^2 -9g_2^2 \lambda_2 +\frac{27}{4}g_2^{4}+\frac{81}{100}g_1^{4}+\frac{9}{10}g_1^2 \Big(-2\lambda_2 +3g_2^2 \Big),\nonumber \\
\beta_{\lambda_{S}} & =3\Big(4\omega_1^2 +4\omega_2^2 +\lambda_{S}^2 \Big),\nonumber \\
\beta_{\omega_1 } & =2\lambda_4 \omega_2 +4\lambda_3 \omega_2 +\omega_1 \Big(2\lambda_1 +4\omega_1 +6y_{t}^2 -\frac{9}{10}g_1^2 -\frac{9}{2}g_2^2 +\lambda_{S}\Big),\nonumber \\
\beta_{\omega_2 } & =2\lambda_2 \omega_2 +2\lambda_4 \omega_1 +4\lambda_3 \omega_1 +4\omega_2^2 -\frac{9}{10}g_1^2 \omega_2 -\frac{9}{2}g_2^2 \omega_2 +\lambda_{S}\omega_2 \nonumber \\
\beta_{\lambda_3 } & =+\frac{27}{100}g_1^{4}+\frac{9}{4}g_2^{4}-9g_2^2 \lambda_3 +2\lambda_1 \lambda_3 +2\lambda_2 \lambda_3 +4\lambda_3^2 -\frac{9}{10}g_1^2 \Big(2\lambda_3 +g_2^2 \Big)+\frac{2}{3}\lambda_1 \lambda_4 +\frac{2}{3}\lambda_2 \lambda_4 +2\lambda_4^2 +2\lambda_5^2 \nonumber \\
 & +\omega_1 \omega_2 +6\lambda_3 y_{t}^2 ,\nonumber \\
\beta_{\lambda_4 } & =4\lambda_4^2 +6\lambda_4 y_{t}^2 +8\lambda_3 \lambda_4 +8\lambda_5^2 -9g_2^2 \lambda_4 +\frac{2}{3}\lambda_1 \lambda_4 +\frac{2}{3}\lambda_2 \lambda_4 +\frac{9}{5}g_1^2 \Big(-\lambda_4 +g_2^2 \Big),\nonumber \\
\beta_{\lambda_5 } & =\lambda_5 \Big(6y_{t}^2 -9g_2^2 +8\lambda_3 +12\lambda_4 +\frac{2}{3}\lambda_1 +\frac{2}{3}\lambda_2 -\frac{9}{5}g_1^2 \Big),\nonumber \\
\beta_{y_{t}} & =\frac{3}{2}y_{t}^{3}+y_{t}\Big(3y_{t}^2 -8g_3^2 -\frac{17}{20}g_1^2 -\frac{9}{4}g_2^2 \Big),
\end{align}
where $g_{i}=(g_1 ,g_2 ,g_3 ) $ represents the SM gauge couplings.

The $\beta $ functions are given at two-loop level in the following subsections by:
\begin{align}
\beta_{g_1 }^{(2)} & =\frac{1}{50}g_1^{3}\Big(180g_2^2 +208g_1^2 +440g_3^2 -85y_{t}^2 \Big)\\
\beta_{g_2 }^{(2)} & =\frac{1}{10}g_2^{3}\Big(120g_3^2 +12g_1^2 -15y_{t}^2 +80g_2^2 \Big)\\
\beta_{g_3 }^{(2)} & =\frac{1}{10}g_3^{3}\Big(11g_1^2 -20y_{t}^2 -260g_3^2 +45g_2^2 \Big)
\end{align}

\begin{align}
\beta_{y_{t}}^{(2)} & =
y_{t}\Big(\frac{1267}{600}g_1^{4}-\frac{9}{20}g_1^2 g_2^2 -\frac{21}{4}g_2^{4}+\frac{19}{15}g_1^2 g_3^2 +9g_2^2 g_3^2 -108g_3^{4}+\frac{1}{6}\lambda_1^2 +\lambda_3^2 +\lambda_3 \lambda_4 +\lambda_4^2 +\frac{3}{2}\lambda_5^2 +\frac{1}{4}\omega_1^2 \Big)\nonumber\\&
-12y_{t}^{5}+y_{t}^{3}\Big([\frac{79}{80}+\frac{8}{5}+\frac{17}{8}+\frac{8}{25}]g_1^2 +\frac{225}{16}g_2^2 +\frac{228}{5}g_3^2 -2\lambda_1 \Big)
\end{align}

\begin{align}
 \beta_{\lambda_5 }^{(2)} & =+\frac{1413}{200}g_1^{4}\lambda_5 +\frac{57}{20}g_1^2 g_2^2 \lambda_5 -\frac{231}{8}g_2^{4}\lambda_5 -\frac{2}{5}g_1^2 \lambda_1 \lambda_5 -\frac{7}{9}\lambda_1^2 \lambda_5 -\frac{2}{5}g_1^2 \lambda_2 \lambda_5 -\frac{7}{9}\lambda_2^2 \lambda_5 \nonumber \\
 & +\frac{48}{5}g_1^2 \lambda_3 \lambda_5 +36g_2^2 \lambda_3 \lambda_5 -\frac{40}{3}\lambda_1 \lambda_3 \lambda_5 -\frac{40}{3}\lambda_2 \lambda_3 \lambda_5 -28\lambda_3^2 \lambda_5 +\frac{72}{5}g_1^2 \lambda_4 \lambda_5 +72g_2^2 \lambda_4 \lambda_5 \nonumber \\
 & -\frac{44}{3}\lambda_1 \lambda_4 \lambda_5 -\frac{44}{3}\lambda_2 \lambda_4 \lambda_5 -76\lambda_3 \lambda_4 \lambda_5 -32\lambda_4^2 \lambda_5 +6\lambda_5^{3}-\frac{1}{2}\lambda_5 \omega_1^2 -4\lambda_5 \omega_1 \omega_2 -\frac{1}{2}\lambda_5 \omega_2^2 \nonumber \\
 & +\frac{1}{4}\Big(16\Big(10g_3^2 -6\lambda_3 -9\lambda_4 -\lambda_1 \Big)+17g_1^2 +45g_2^2 \Big)\lambda_5 y_{t}^2 -\frac{3}{2}\lambda_5 y_{t}^{4} 
\end{align}

\begin{align}
\beta_{\lambda_{S}}^{(2)} & =-20\lambda_{S}\Big(\omega_1^2 +\omega_2^2 \Big)-48\Big(\omega_1^{3}+\omega_2^{3}\Big)+72g_2^2 \Big(\omega_1^2 +\omega_2^2 \Big)-72\omega_1^2 y_{t}^2 -\frac{17}{3}\lambda_{S}^{3}+\frac{72}{5}g_1^2 \Big(\omega_1^2 +\omega_2^2 \Big) 
\end{align}

\begin{align}
\beta_{\omega_1 }^{(2)} & =+\frac{1737}{400}g_1^{4}\omega_1 +\frac{9}{8}g_1^2 g_2^2 \omega_1 -\frac{123}{16}g_2^{4}\omega_1 +\frac{12}{5}g_1^2 \lambda_1 \omega_1 +12g_2^2 \lambda_1 \omega_1 -\frac{5}{3}\lambda_1^2 \omega_1 -2\lambda_3^2 \omega_1 -2\lambda_3 \lambda_4 \omega_1 \nonumber \\
 & -2\lambda_4^2 \omega_1 -3\lambda_5^2 \omega_1 -\frac{5}{6}\lambda_{S}^2 \omega_1 +\frac{3}{5}g_1^2 \omega_1^2 +3g_2^2 \omega_1^2 -12\lambda_1 \omega_1^2 -6\lambda_{S}\omega_1^2 -\frac{21}{2}\omega_1^{3}+\frac{9}{10}g_1^{4}\omega_2 \nonumber \\
 & +\frac{15}{2}g_2^{4}\omega_2 +\frac{24}{5}g_1^2 \lambda_3 \omega_2 +24g_2^2 \lambda_3 \omega_2 -8\lambda_3^2 \omega_2 +\frac{12}{5}g_1^2 \lambda_4 \omega_2 +12g_2^2 \lambda_4 \omega_2 -8\lambda_3 \lambda_4 \omega_2 \nonumber \\
 & -8\lambda_4^2 \omega_2 -12\lambda_5^2 \omega_2 -16\lambda_3 \omega_1 \omega_2 -8\lambda_4 \omega_1 \omega_2 -8\lambda_3 \omega_2^2 -4\lambda_4 \omega_2^2 -2\omega_1 \omega_2^2 \nonumber \\
 & +\frac{1}{4}\Big(160g_3^2 +17g_1^2 +45g_2^2 -48\lambda_1 -48\omega_1 \Big)\omega_1 y_{t}^2 -\frac{27}{2}\omega_1 y_{t}^{4} 
\end{align}

\begin{align}
\beta_{\lambda_1 }^{(2)} & =-\frac{10611}{1000}g_1^{6}-\frac{5157}{200}g_1^{4}g_2^2 -\frac{909}{40}g_1^2 g_2^{4}+\frac{873}{8}g_2^{6}+\frac{1953}{200}g_1^{4}\lambda_1 +\frac{117}{20}g_1^2 g_2^2 \lambda_1 -\frac{51}{8}g_2^{4}\lambda_1 +\frac{18}{5}g_1^2 \lambda_1^2 \nonumber \\
 & +18g_2^2 \lambda_1^2 -\frac{26}{3}\lambda_1^{3}+\frac{27}{5}g_1^{4}\lambda_3 +45g_2^{4}\lambda_3 +\frac{72}{5}g_1^2 \lambda_3^2 +72g_2^2 \lambda_3^2 -20\lambda_1 \lambda_3^2 -48\lambda_3^{3}+\frac{27}{10}g_1^{4}\lambda_4 \nonumber \\
 & +9g_1^2 g_2^2 \lambda_4 +\frac{45}{2}g_2^{4}\lambda_4 +\frac{72}{5}g_1^2 \lambda_3 \lambda_4 +72g_2^2 \lambda_3 \lambda_4 -20\lambda_1 \lambda_3 \lambda_4 -72\lambda_3^2 \lambda_4 +\frac{36}{5}g_1^2 \lambda_4^2 \nonumber \\
 & +18g_2^2 \lambda_4^2 -12\lambda_1 \lambda_4^2 -96\lambda_3 \lambda_4^2 -36\lambda_4^{3}-\frac{18}{5}g_1^2 \lambda_5^2 -14\lambda_1 \lambda_5^2 -120\lambda_3 \lambda_5^2 -132\lambda_4 \lambda_5^2 \nonumber \\& -5\lambda_1 \omega_1^2 -12\omega_1^{3}-\frac{1}{50}\Big(25\Big(-160g_3^2 \lambda_1 +27g_2^{4}-45g_2^2 \lambda_1 +48\lambda_1^2 \Big)+513g_1^{4}\nonumber\\&-5g_1^2 \Big(378g_2^2 +85\lambda_1 \Big)\Big)y_{t}^2 -\frac{3}{5}\Big(16g_1^2 +320g_3^2 +5\lambda_1 \Big)y_{t}^{4}+180y_{t}^{6} 
\end{align}

\begin{align}
\beta_{\lambda_4 }^{(2)} & =-\frac{657}{50}g_1^{4}g_2^2 -\frac{42}{5}g_1^2 g_2^{4}+g_1^2 g_2^2 \lambda_1 +g_1^2 g_2^2 \lambda_2 +\frac{6}{5}g_1^2 g_2^2 \lambda_3 +\frac{1413}{200}g_1^{4}\lambda_4 +\frac{153}{20}g_1^2 g_2^2 \lambda_4 \nonumber \\
 & -\frac{231}{8}g_2^{4}\lambda_4 +\frac{4}{5}g_1^2 \lambda_1 \lambda_4 -\frac{7}{9}\lambda_1^2 \lambda_4 +\frac{4}{5}g_1^2 \lambda_2 \lambda_4 -\frac{7}{9}\lambda_2^2 \lambda_4 +\frac{12}{5}g_1^2 \lambda_3 \lambda_4 +36g_2^2 \lambda_3 \lambda_4 \nonumber \\
 & -\frac{40}{3}\lambda_1 \lambda_3 \lambda_4 -\frac{40}{3}\lambda_2 \lambda_3 \lambda_4 -28\lambda_3^2 \lambda_4 +\frac{24}{5}g_1^2 \lambda_4^2 +18g_2^2 \lambda_4^2 -\frac{20}{3}\lambda_1 \lambda_4^2 -\frac{20}{3}\lambda_2 \lambda_4^2 -28\lambda_3 \lambda_4^2 \nonumber \\
 & +\frac{48}{5}g_1^2 \lambda_5^2 +54g_2^2 \lambda_5^2 -8\lambda_1 \lambda_5^2 -8\lambda_2 \lambda_5^2 -48\lambda_3 \lambda_5^2 -26\lambda_4 \lambda_5^2 -\frac{1}{2}\lambda_4 \omega_1^2 -4\lambda_4 \omega_1 \omega_2 \nonumber \\
 & -\frac{1}{2}\lambda_4 \omega_2^2 +\Big(40g_3^2 \lambda_4 -4\Big(3\lambda_4^2 +6\lambda_3 \lambda_4 +6\lambda_5^2 +\lambda_1 \lambda_4 \Big)+\frac{45}{4}g_2^2 \lambda_4 +g_1^2 \Big(\frac{17}{4}\lambda_4 +\frac{63}{5}g_2^2 \Big)\Big)y_{t}^2 \nonumber \\
 & -\frac{27}{2}\lambda_4 y_{t}^{4}
\end{align}

\begin{align}
\beta_{\omega_2 }^{(2)} & =+\frac{15}{2}g_2^{4}\omega_1 +24g_2^2 \lambda_3 \omega_1 -8\lambda_3^2 \omega_1 +12g_2^2 \lambda_4 \omega_1 -8\lambda_3 \lambda_4 \omega_1 -8\lambda_4^2 \omega_1 -12\lambda_5^2 \omega_1 -8\lambda_3 \omega_1^2 \nonumber \\
 & -4\lambda_4 \omega_1^2 -\frac{123}{16}g_2^{4}\omega_2 +12g_2^2 \lambda_2 \omega_2 -\frac{5}{3}\lambda_2^2 \omega_2 -2\lambda_3^2 \omega_2 -2\lambda_3 \lambda_4 \omega_2 -2\lambda_4^2 \omega_2 -3\lambda_5^2 \omega_2 \nonumber \\
 & -\frac{5}{6}\lambda_{S}^2 \omega_2 -16\lambda_3 \omega_1 \omega_2 -8\lambda_4 \omega_1 \omega_2 -2\omega_1^2 \omega_2 +3g_2^2 \omega_2^2 -12\lambda_2 \omega_2^2 -6\lambda_{S}\omega_2^2 -\frac{21}{2}\omega_2^{3}\nonumber \\
 & +\frac{9}{400}g_1^{4}\Big(193\omega_2 +40\omega_1 \Big)+\frac{3}{40}g_1^2 \Big(32\lambda_4 \omega_1 +64\lambda_3 \omega_1 +\omega_2 \Big(15g_2^2 +32\lambda_2 +8\omega_2 \Big)\Big)\nonumber \\&-12\Big(2\lambda_3 +\lambda_4 \Big)\omega_1 y_{t}^2 
\end{align}

\begin{align}
 \beta_{\lambda_3 }^{(2)} & =-\frac{3537}{1000}g_1^{6}+\frac{909}{200}g_1^{4}g_2^2 +\frac{33}{40}g_1^2 g_2^{4}+\frac{291}{8}g_2^{6}+\frac{9}{20}g_1^{4}\lambda_1 -\frac{1}{2}g_1^2 g_2^2 \lambda_1 +\frac{15}{4}g_2^{4}\lambda_1 +\frac{9}{20}g_1^{4}\lambda_2 \nonumber \\
 & -\frac{1}{2}g_1^2 g_2^2 \lambda_2 +\frac{15}{4}g_2^{4}\lambda_2 +\frac{1773}{200}g_1^{4}\lambda_3 +\frac{33}{20}g_1^2 g_2^2 \lambda_3 -\frac{111}{8}g_2^{4}\lambda_3 +\frac{12}{5}g_1^2 \lambda_1 \lambda_3 +12g_2^2 \lambda_1 \lambda_3 \nonumber \\
 & -\frac{5}{3}\lambda_1^2 \lambda_3 +\frac{12}{5}g_1^2 \lambda_2 \lambda_3 +12g_2^2 \lambda_2 \lambda_3 -\frac{5}{3}\lambda_2^2 \lambda_3 +\frac{6}{5}g_1^2 \lambda_3^2 +6g_2^2 \lambda_3^2 -12\lambda_1 \lambda_3^2 -12\lambda_2 \lambda_3^2 \nonumber \\
 & -12\lambda_3^{3}+\frac{9}{10}g_1^{4}\lambda_4 -\frac{9}{5}g_1^2 g_2^2 \lambda_4 +\frac{15}{2}g_2^{4}\lambda_4 +\frac{4}{5}g_1^2 \lambda_1 \lambda_4 +6g_2^2 \lambda_1 \lambda_4 -\frac{4}{9}\lambda_1^2 \lambda_4 +\frac{4}{5}g_1^2 \lambda_2 \lambda_4 \nonumber \\
 & +6g_2^2 \lambda_2 \lambda_4 -\frac{4}{9}\lambda_2^2 \lambda_4 -12g_2^2 \lambda_3 \lambda_4 -\frac{16}{3}\lambda_1 \lambda_3 \lambda_4 -\frac{16}{3}\lambda_2 \lambda_3 \lambda_4 -4\lambda_3^2 \lambda_4 -\frac{6}{5}g_1^2 \lambda_4^2 +6g_2^2 \lambda_4^2 \nonumber \\
 & -\frac{14}{3}\lambda_1 \lambda_4^2 -\frac{14}{3}\lambda_2 \lambda_4^2 -16\lambda_3 \lambda_4^2 -12\lambda_4^{3}+\frac{12}{5}g_1^2 \lambda_5^2 -6\lambda_1 \lambda_5^2 -6\lambda_2 \lambda_5^2 -18\lambda_3 \lambda_5^2 -44\lambda_4 \lambda_5^2 \nonumber \\
 & -\frac{1}{2}\lambda_3 \omega_1^2 -4\lambda_3 \omega_1 \omega_2 -2\omega_1^2 \omega_2 -\frac{1}{2}\lambda_3 \omega_2^2 -2\omega_1 \omega_2^2 +5g_1^2 \Big(126g_2^2 -85\lambda_3 \Big)\Big)y_{t}^2 -\frac{27}{2}\lambda_3 y_{t}^{4} \nonumber \\
 & -\frac{1}{100}\Big(171g_1^{4}+25\Big(-45g_2^2 \lambda_3 +8\Big(-20g_3^2 \lambda_3 +2\lambda_1 \Big(3\lambda_3 +\lambda_4 \Big)+3\Big(2\lambda_3^2 +\lambda_4^2 +\lambda_5^2 \Big)\Big)+9g_2^{4}\Big) 
\end{align}

\begin{align}
\beta_{\lambda_2 }^{(2)} & =-\frac{10611}{1000}g_1^{6}+\frac{873}{8}g_2^{6}-\frac{51}{8}g_2^{4}\lambda_2 +18g_2^2 \lambda_2^2 -\frac{26}{3}\lambda_2^{3}+45g_2^{4}\lambda_3 +72g_2^2 \lambda_3^2 -20\lambda_2 \lambda_3^2 -48\lambda_3^{3}\nonumber \\
 &+\frac{45}{2}g_2^{4}\lambda_4 +72g_2^2 \lambda_3 \lambda_4 -20\lambda_2 \lambda_3 \lambda_4 -72\lambda_3^2 \lambda_4 +18g_2^2 \lambda_4^2 -12\lambda_2 \lambda_4^2 -96\lambda_3 \lambda_4^2 -36\lambda_4^{3}\nonumber \\
 & -\frac{9}{200}g_1^{4}\Big(-217\lambda_2 +573g_2^2 -60\Big(2\lambda_3 +\lambda_4 \Big)\Big)-14\lambda_2 \lambda_5^2 -120\lambda_3 \lambda_5^2 -132\lambda_4 \lambda_5^2 \nonumber \\
 & -\frac{9}{40}g_1^2 \Big(101g_2^{4}-16\Big(2\lambda_4^2 +4\lambda_3^2 +4\lambda_3 \lambda_4 -\lambda_5^2 +\lambda_2^2 \Big)-2g_2^2 \Big(13\lambda_2 +20\lambda_4 \Big)\Big)-5\lambda_2 \omega_2^2 \nonumber \\
 & -12\omega_2^{3}-36\Big(2\lambda_3^2 +2\lambda_3 \lambda_4 +\lambda_4^2 +\lambda_5^2 \Big)y_{t}^2 
\end{align}

\end{document}